\documentclass[12pt]{article}   
\pdfoutput=1	
\usepackage{geometry,bbm}   
\usepackage[toc,page]{appendix}             		
\usepackage{graphicx}
\usepackage[usenames,dvipsnames]{color}
\usepackage{slashed}					
\usepackage{amssymb,amsmath}
\usepackage{hyperref}
\usepackage{mathtools}
\usepackage{multirow}
\usepackage[table]{xcolor}
\usepackage{subfigure}

\newcommand{\nn}{\nonumber}

\def\){\right)}
\def\({\left( }
\def\]{\right] }
\def\[{\left[ }

\newcommand{\ea}{\end{align*}}
\newcommand{\ba}{\begin{align*}}
\newcommand{\bi}{\begin{itemize}}
\newcommand{\ei}{\end{itemize}}

\newcommand{\ben}{\begin{enumerate}}
\newcommand{\een}{\end{enumerate}}

\newcommand{\Oc}{\mathcal{O}}

\def\){\right)}
\def\({\left( }
\def\]{\right] }
\def\[{\left[ }
\newcommand{\la}{\langle}
\newcommand{\ra}{\rangle}

\def\NO{\nonumber}

\newcommand{\be}{\begin{equation}}
\newcommand{\ee}{\end{equation}}

\def\bea{\begin{eqnarray}}
\def\eea{\end{eqnarray}}

\def\bal#1\eal{\begin{align}#1\end{align}}

\def\bald{\begin{aligned}}
\def\eald{\end{aligned}}

\def\beqx{\begin{displaymath}}
\def\eeqx{\end{displaymath}}

\newcommand{\bmat}{\left(\begin{array}}
\newcommand{\emat}{\end{array}\right)}

\def\b{\beta}

\def\d{\delta}

\def\f{\phi}
\def\g{\gamma}

\def\k{\kappa}
\def\l{\lambda}
\def\m{\mu}
\def\n{\nu}

\def\p{\pi}

\def\r{\rho}

\def\x{\xi}

\def\F{\Phi}
\def\G{\Gamma}

\def\ba{\bbalpha}

\def\cn{{\cal N}}
\def\co{{\cal O}}

\def\bo{{\raise-.3ex\hbox{\large$\Box$}}}               
\def\pa{\partial}                                       
\def\de{\nabla}                                         
\def\face{{\raise.2ex\hbox{$\displaystyle \bigodot$}\mskip-2.2mu \llap {$\ddot
        \smile$}}}                                   
\def\>{\rangle}                                      
\def\<{\langle}                                      

\def\leftrightarrowfill{$\mathsurround=0pt \mathord\leftarrow \mkern-6mu
        \cleaders\hbox{$\mkern-2mu \mathord- \mkern-2mu$}\hfill
        \mkern-6mu \mathord\rightarrow$}        
\def\dvec#1{\vbox{\ialign{##\crcr
        \leftrightarrowfill\crcr\noalign{\kern-1pt\nointerlineskip}
        $\hfil\displaystyle{#1}\hfil$\crcr}}}           
\def\Re{{\rm Re\,}}                                     
\def\Im{{\rm Im\,}}                                     

\def\-{\hphantom{-}}

\numberwithin{equation}{section}
\begin{document}

\begin{titlepage}

\begin{flushright}
SISSA 43/2017/FISI
\end{flushright}
\bigskip
\def\thefootnote{\fnsymbol{footnote}}

\begin{center}
{\LARGE
{\bf
On non-supersymmetric conformal \\ \vskip 15pt manifolds: field theory and holography}}
\end{center}

\bigskip
\begin{center}
{\large
Vladimir Bashmakov$^{1}$, Matteo Bertolini$^{1,2}$, Himanshu Raj$^1$}

\end{center}

\renewcommand{\thefootnote}{\arabic{footnote}}

\begin{center}
\vspace{0.2cm}
$^1$ {SISSA and INFN - 
Via Bonomea 265; I 34136 Trieste, Italy\\}
$^2$ {ICTP - 
Strada Costiera 11; I 34014 Trieste, Italy\\}

\vskip 5pt
{\texttt{bertmat,hraj,vbashmakov @sissa.it}}

\end{center}

\vskip 5pt
\noindent
\begin{center} {\bf Abstract} \end{center}
\noindent
We discuss the constraints that a conformal field theory should enjoy to admit exactly marginal deformations, {\it i.e.} to be part of a conformal manifold. In particular, using tools from conformal perturbation theory, we derive a sum rule from which one can extract restrictions on the spectrum of low spin operators and on the behavior of OPE coefficients involving nearly marginal operators. We then focus on conformal field theories admitting a gravity dual description, and as such a large-$N$ expansion. We discuss the relation between conformal perturbation theory and loop expansion in the bulk, and show how such connection could help in the search for conformal manifolds beyond the planar limit. Our results do not rely on supersymmetry, and therefore apply also outside the realm of superconformal field theories. 

\vspace{1.6 cm}
\vfill

\end{titlepage}

\newpage
\tableofcontents

\section{Introduction}
\label{Introduction}

It has been known since many years that there exist families of superconformal field theories (SCFTs) connected by exactly marginal deformations \cite{Leigh:1995ep} (see, e.g., \cite{Kol:2002zt,Benvenuti:2005wi,Green:2010da,Kol:2010ub} for generalizations). The corresponding exactly marginal couplings parametrize what is known as the conformal manifold. 

An obvious question is whether conformal manifolds can exist even in absence of supersymmetry. Unless there exist some other underlying extended symmetries, general arguments suggest this to be hardly possible. Upon deforming a conformal field theory (CFT) as
\be
\label{conman1}
S_{CFT} \rightarrow S_{CFT} + g \int d^dx \,{\cal O}~,
\ee
where ${\cal O}$ is a scalar primary of the CFT with scaling dimension $\Delta_{\cal O}=d$, a $\beta$ function for the coupling $g$ is induced, at the quantum level. The existence of a conformal manifold requires $\beta(g)=0$ and it is hard to believe this to be possible without supersymmetry, which, in some circumstances \cite{Leigh:1995ep}, can in fact protect ${\cal O}$ from acquiring an anomalous dimension. Moreover, the deformation triggered by the coupling $g$ could also generate new couplings at the quantum level, and the corresponding $\beta$ functions should also be set to zero, if we were to preserve conformal invariance. Therefore,  constraints look rather tight.

One could wonder whether there exist some consistency constraints that forbid a non-supersymmetric conformal manifold to exist, to start with. While we are not aware of any no-go theorem, the following simple argument shows that non-supersymmetric conformal manifolds can be consistent at least with unitarity and crossing symmetry.

As we will review later, the requirement of vanishing $\b$ functions imposes stringent constraints on the CFT data, but only regarding operators with integer spins. One can then take any of the known SCFTs belonging to a conformal manifold and truncate the spectrum of operators, excluding all operators with half-integer spins, while leaving CFT data of integer-spin operators unmodified. This is consistent, because half-integer spin operators cannot appear in the OPE of two integer spin operators. The operator algebra one ends up with is crossing-symmetric because initially it was, and also the truncated Hilbert space does not contain any negative-norm states, because the original one did not, consistently with unitarity. CFT data still obey the $\b$-function constraints, because the original theory had a conformal manifold by assumption. And, finally, the resulting operator algebra does not form a representation of the supersymmetry algebra, because it contains only integer spin operators. This might suggest it to be simple, eventually, to construct non-supersymmetric CFTs living on a conformal manifold. In fact, unitarity and crossing symmetry are necessary but not sufficient conditions to get a consistent theory.\footnote{We thank Alexander Zhiboedov for a discussion on this point.} For instance, there are further conditions coming from modular invariance in two dimensions or, more generally, by requiring the consistency of the CFT at finite temperature in any number of dimensions, see, {\it e.g.}, \cite{ElShowk:2011ag}. This is why the truncation described above does not allow for getting non-supersymmetric conformal manifolds for free. The truncated operator algebra might not form a consistent CFT, eventually. It will be interesting to investigate this issue further. In this work, we will just assume that non-supersymmetric conformal manifolds can exist, and elaborate upon the corresponding constraints.

The very possibility for a conformal manifold to exist requires the presence of one (or more) marginal scalar operator in the undeformed CFT, an operator ${\cal O}$ with scaling dimension $\Delta_{\cal O} = d$. This implies that $\beta(g)$ vanishes, at tree level in $g$. We want to investigate which further conditions the requirement of vanishing $\beta$ function at the quantum level imposes on the CFT. To put things the simplest, we will focus on one-dimensional conformal manifolds, described by deformations like \eqref{conman1}.

In section 2, using conformal perturbation theory, we start by reviewing the conditions that the vanishing of the $\b$ function up to two-loop order imposes on the OPE of the operator ${\cal O}$. Then, using also recent numerical bootstrap results, we show what other information on the spectrum of low dimension operators other than ${\cal O}$,  can be extracted. This includes, in particular, the dependence on scaling dimension of OPE coefficients involving nearly marginal operators $\Delta \sim d$, as well as a prediction on the content of low spin operators in the spectrum of the CFT.

In section 3, we focus on CFTs admitting a gravity dual description. First, we discuss the relation between conformal perturbation theory and the $1/N$ expansion, and the role that Witten diagrams play in this matter. Then, focusing on a toy-model, we investigate under which conditions a conformal manifold existing at leading order in $1/N$, can survive at non-planar level, and show that, even in absence of supersymmetry, this is a non-empty set. On the way, we also provide a nice AdS/CFT consistency check regarding non-supersymmetric AdS (in)stability and CFTs RG flows. 

Section 4, which is our last section, contains a discussion on models with richer dynamics, and an outlook on what one can do next using our results.


\section{Constraints from conformal perturbation theory}
 \label{cpt}

Given a CFT and a deformation as that in eq.~\eqref{conman1}, one expects that a $\beta$ function for the coupling $g$ is generated and that conformal invariance is lost. The $\beta$ function reads
\be
\label{betafull}
\beta(g) = \beta_1\, g^2 +  \beta_2 \, g^3 + \dots ~.
\ee
Loop coefficients are expected to depend on the data of undeformed CFT. In order to find such dependence a perturbative analysis can be conveniently done in the context of conformal perturbation theory (CPT) \cite{Cardy:1987vr}. 

One can extract the $\beta$ function by considering cleverly chosen physical observables and demand them to be UV-cutoff independent. Following  \cite{Komargodski:2016auf} (see also \cite{Behan:2017dwr}), we consider the overlap 
\be
\label{overlap1}
\langle {\cal O}(\infty)|0\rangle_{g,V} 
\ee
where 
${\cal O}(\infty) = \lim_{x\rightarrow \infty} x^{2d} {\cal O}(x)$, while $|0\rangle_{g,V} = e^{ g \int_V d^dx \,{\cal O}(x)} |0\rangle$ is the state obtained by deforming the theory by  \eqref{conman1} in a finite region around the origin. 
The choice of a finite volume $V$ allows one to get rid of IR divergences, while not affecting the UV behavior we are interested in. Expanding \eqref{overlap1} in $g$ one gets a perturbative expansion in terms of integrals of $n$-point functions of ${\cal O}$. These are generically plagued by logarithmic divergences, which can be absorbed by demanding that the coupling $g$ runs with scale $\mu$ in a way that the final result is $\mu$-independent. This, in turns, lets one extract the $\beta$ function. 

Proceeding this way one gets for the $\beta$ function at two loops (which to this order is universal, hence independent of the renormalization scheme) the following expressions
\bea
\label{1lg}
\beta_1 &=& - \frac 12 \, S_{d-1} C_{{\cal O}{\cal O}{\cal O}}  \\
\label{2lg}
 \beta_2 &=&-\frac 16 \, S_{d-1} \int d^dx \Big[ \langle {\cal O}(0)  {\cal O}(x)  {\cal O}(e)  {\cal O}(\infty)\rangle_c 
 - \sum_{\Phi}\frac 12  C_{{\cal O O}  \Phi}^2 
 \left( \frac{1}{x^d (x -e)^d} +  \frac{1}{x^d} +  \frac{1}{(x -e)^d} \right) \nn \\
 && -  \sum_{\Psi} C_{{\cal O O}  \Psi}^2 
 \left( \frac{1}{x^{2d-\Delta_{\Psi}}} +  \frac{1}{(x-e)^{2d-\Delta_{\Psi}}} +  x^{-\Delta_{\Psi}} \right)\Big]~,
\eea
where $S_{d-1}$ is the volume of the ($d-1$)-dimensional unit sphere, $e$ is a unit vector in some fixed direction and the subscript $c$ in the four-point function refers to the connected contribution. Sums are over marginal operators $\Phi$ and relevant operators $\Psi$ appearing in the $\cal O O$ OPE. In principle, one can go to higher orders in $g$. In particular, marginality of ${\cal O}$ at order $O(g^{n-1})$  would require the vanishing of logarithmic divergences of an integral  in $d^dx_1 \cdots d^dx_{n-3}$ of the $n$-point function $\langle {\cal O} \dots {\cal O}\rangle $.

The deformation \eqref{conman1} does not cause the running of $g$, only. In general, any coupling $g_\Phi$ dual to a marginal operator $\Phi$ appearing in the OPE of ${\cal O}(x){\cal O}(0)$ will start running, due to quantum effects.\footnote{Runnings are also induced for relevant operators appearing in the OPE. However, these effects are associated to power-law divergences and can be reabsorbed by local counter-terms. This is  equivalent to be at a fixed point, to ${\cal O}(g^2)$ order, of the corresponding $\b$ functions $\b(g_\Psi)$ \cite{Cardy:1987vr}.} Following the same procedure described above, one gets the following contribution at order $g^2$ to $\beta(g_\varPhi)$
\be
\label{1lgp}
\beta(g_\varPhi) \supset - \frac 12 \, S_{d-1} C_{{\cal O}{\cal O}\varPhi} \, g^2 ~.
\ee
Therefore, at one loop in CPT, the persistence of a conformal manifold under the deformation \eqref{conman1} implies the following constraints on the OPE coefficients of the CFT
\be
\label{1lc}
C_{{\cal O}{\cal O}\varPhi} = 0 ~~,~~\forall \,\varPhi~~\mbox{such~that}~~\Delta_\varPhi = d~.
\ee
Taking into account the above constraint, eq.~\eqref{2lg} simplifies and we get the following condition at two-loops, eventually
\be
\label{2lc}
\int d^dx \left[\, \langle {\cal O}(0)  {\cal O}(x)  {\cal O}(e)  {\cal O}(\infty)\rangle_c-  \sum_{\Psi} C_{{\cal O O}  \Psi}^2 
 \left( \frac{1}{x^{2d-\Delta_{\Psi}}} +  \frac{1}{(x-e)^{2d-\Delta_{\Psi}}} +  x^{-\Delta_{\Psi}} \right)\right] = 0~.
\ee
Eqs.~\eqref{1lc} and \eqref{2lc} are the two constraints the existence of a conformal manifold under the deformation \eqref{conman1} imposes on the CFT at two-loop order in CPT.\footnote{One can obtain similar expressions for two-loop $\b$ function of other marginal operators, if there are any, and get additional constraints.}

\subsection{Two-loop constraint and integrated conformal blocks}
\label{2l}

One can try to translate the constraint \eqref{2lc} into a sum rule in terms of conformal blocks, which can provide, in turn, constraints on the CFT data. 

Let us first rewrite \eqref{2lc} as an integral of the full four-point function, that is
\begin{eqnarray}
\label{2lcbis}
\int d^dx \, &\Big(\langle {\cal O}(0)  {\cal O}(x)  {\cal O}(e)  {\cal O}(\infty)\rangle - \frac{1}{x^{2d}} - \frac{1}{(x-e)^{2d}} - 1 -\nonumber\\
&\sum_{\Psi} C_{{\cal O O}  \Psi}^2 
 \left( \frac{1}{x^{2d-\Delta_{\Psi}}} +  \frac{1}{(x-e)^{2d-\Delta_{\Psi}}} +  x^{-\Delta_{\Psi}} \right)\Big)= 0~.
\end{eqnarray}
The integrand above is axial-symmetric, hence the integration can be seen as an integration over a two-plane $(z , \bar z)$ containing the unit vector $e$, followed by  integration over a $(d-2)$-dimensional sphere, whose coordinates the integrand does not depend on. So, for the integration measure, we get 
\begin{equation}
d^dx\rightarrow\frac{\pi^{\frac{d-1}{2}}}{2\Gamma\(\frac{d-1}{2}\)}d^2z\, \(\frac{z-\bar{z}}{2i}\)^{d-2}~.
\end{equation}
Notice that the integrand together with the measure is inversion-invariant. Therefore, instead of integrating over the whole $R^d$, one can integrate over a unit disk, $B_{r=1}(0)=\{z\in C \,,\, |z|\leq1\}$, where the coordinate $z$ is chosen such that $x=e$ corresponds to $z=1$. 

The integrand in eq.~\eqref{2lcbis} is expected to be a singularity-free function, but among the terms coming with a minus sign, there are some which have manifest singularities. Hence, they must be compensated by the corresponding singularities of the four-point function. 
Due to divergences both at $z=0$ and $z=1$, one cannot use just one OPE channel. However, it turns out that one can reduce the integration domain to a fundamental one  \cite{Gillioz:2016jnn}, for which a single channel suffices. The integral \eqref{2lcbis} is invariant under transformations generated by $z\rightarrow 1/z$ and $z\rightarrow 1-z$ and complex conjugation. Hence, choosing one of the following domains 
\bea
\label{domains}
&&D_1=\{z\in C|~|1-z|^2<1, ~ \Re(z)<1/2, ~ \Im(z)>0\}  \nn \\
&&D_2=\{z\in C|~|1-z|^2<1, ~ \Re(z)<1/2, ~ \Im(z)<0\} \nn \\
&&D_3=\{z\in C|~|1-z|^2>1, ~ |z|^2<1, ~ \Im(z)>0\} \nn \\
&&D_4=\{z\in C|~|1-z|^2>1, ~ |z|^2<1, ~ \Im(z)<0\}~,
\eea
one can use $s$-channel OPE only.
\begin{figure}[ht]
\begin{center}
\includegraphics[height=0.25\textheight]{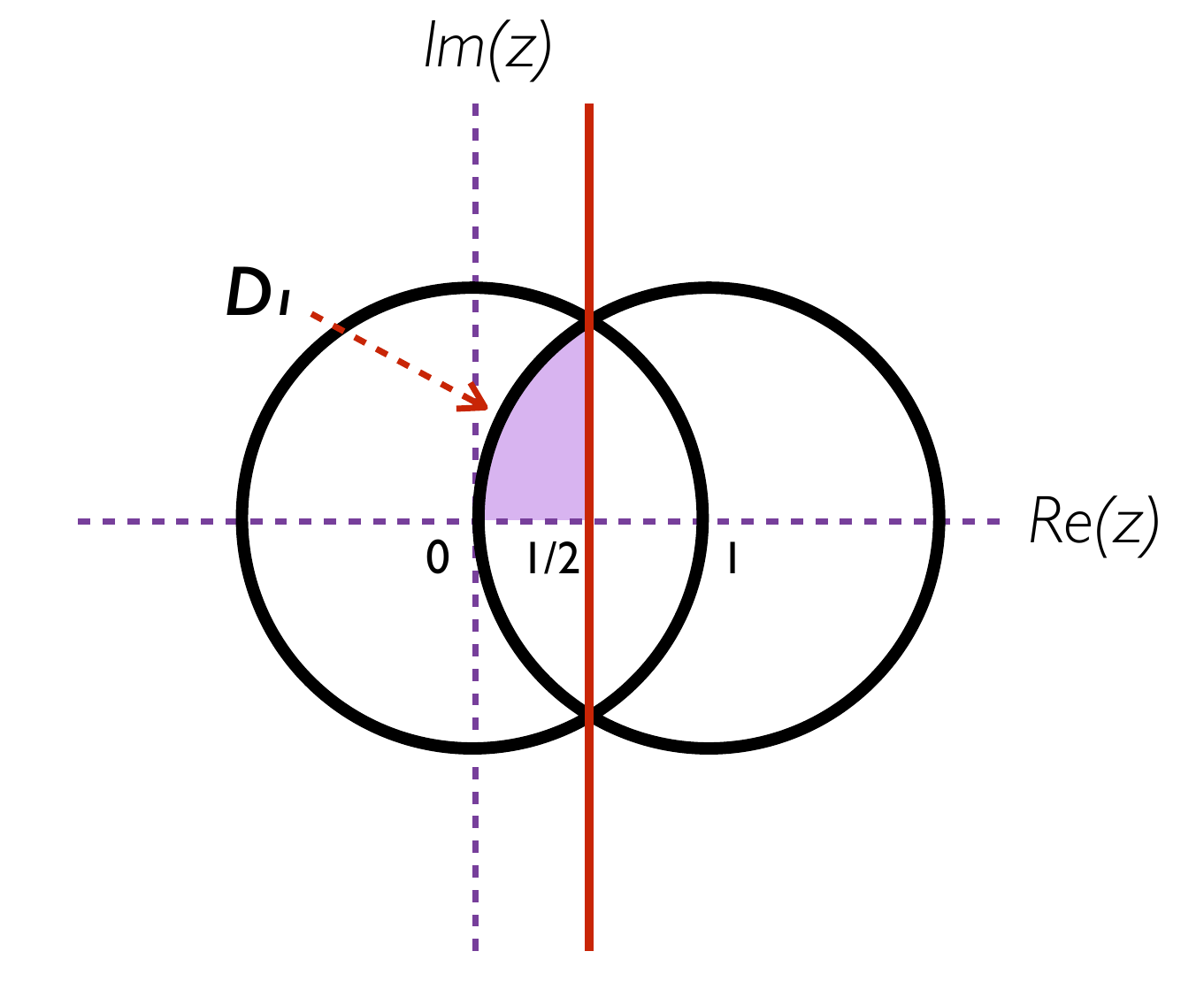}
\caption{Integration in the $(z, \bar z)$ plane. The fundamental domain $D_1$ is the violet region. The regions $D_2, D_3$ and $D_4$ are defined in \eqref{domains} and are easily recognizable in the figure. }
\label{FundDom}
\end{center}
\end{figure}
For the sake of computational convenience we will not do the minimal choice, but use the union of all four domains, $D=D_1\cup D_2\cup D_3\cup D_4$. Using $s$-channel OPE, we get 
\begin{equation}
\<\Oc(0)\Oc(x)\Oc(e)\Oc(\infty)\>=\frac{\sum_{\Oc'}C_{\Oc\Oc\Oc'}^2g_{\Delta_{\Oc'}, \,l_{\Oc'}}}{x^{2d}}~,
\end{equation}
where $g_{\Delta_{\Oc'},\,l_{\Oc'}}$ are conformal blocks corresponding to the exchange of an operator $\Oc'$ with dimension $\Delta_{\Oc'}$ and spin $l_{\Oc'}$ (with $l_{\Oc'}$ even, as in the OPE of two identical scalars only operators with even spin appear). The identity operator contribution cancels the $1/x^{2d}$ divergent contribution in eq.~\eqref{2lcbis}. 

Let us now define the following quantities
\begin{eqnarray}
\label{intcb1}
G_{\Delta_{\Oc'}, \,l_{\Oc'}}&=&\frac{\pi^{\frac{d-1}{2}}}{2\Gamma\(\frac{d-1}{2}\)} \int_{D}d^2z\, \(\frac{z-\bar{z}}{2i}\)^{d-2}\frac{g_{\Delta_{\Oc'}, \,l_{\Oc'}}(z,\bar{z})}{|z|^{2d}}, ~ \Delta>d~, \\
\label{intcb2}
G_{\Delta_{\Oc'}, \,0}&=&\frac{\pi^{\frac{d-1}{2}}}{2\Gamma\(\frac{d-1}{2}\)} \int_{D}d^2z\, \(\frac{z-\bar{z}}{2i}\)^{d-2}\Big(\frac{g_{\Delta_{\Oc'}, \,0}(z,\bar{z})}{|z|^{2d}}  -  \frac{1}{|z|^{2d-\Delta}} -  \frac{1}{|1-z|^{2d-\Delta}} -  |z|^{-\Delta}\Big),  ~\Delta<d~, \nonumber \\ \label{Icbrel1}\\
A&=&\frac{\pi^{\frac{d-1}{2}}}{2\Gamma\(\frac{d-1}{2}\)}~\int_{D}d^2z\, \(\frac{z-\bar{z}}{2i}\)^{d-2}~\(\frac{1}{|1-z|^{2d}}+1\)~,
\end{eqnarray}
where $G_{\Delta_{\Oc'}, \,l_{\Oc'}}$ are {\it integrated} conformal blocks (note that, for $\Delta < d$, that is eq.~\eqref{intcb2}, only scalar operators are above the unitarity bound) and $A$ is a positive, dimension-dependent number, which in, {\it e.g.}, $d=4$ dimensions reads 
\be
A=\frac{\pi}{24}\(9\sqrt{3}+16\pi\)~.
\label{A4&2}
\ee
Using all above definitions, eq. \eqref{2lcbis} can be rewritten as the following sum rule
\begin{equation}
\label{sumrule1}
\sum_{\Oc'} C_{\Oc\Oc\Oc'}^2G_{\Delta_{\Oc'}, \,l_{\Oc'}}=A~.
\end{equation}
Note that now  the contribution of the identity operator is excluded from the sum. 

\begin{figure}[ht!]\label{figd=2}
     \begin{center}
        \subfigure[$l=0, ~\Delta<4$ (relevant scalar) ]{%
            \label{fig:l=0rel}
            \includegraphics[width=0.4\textwidth]{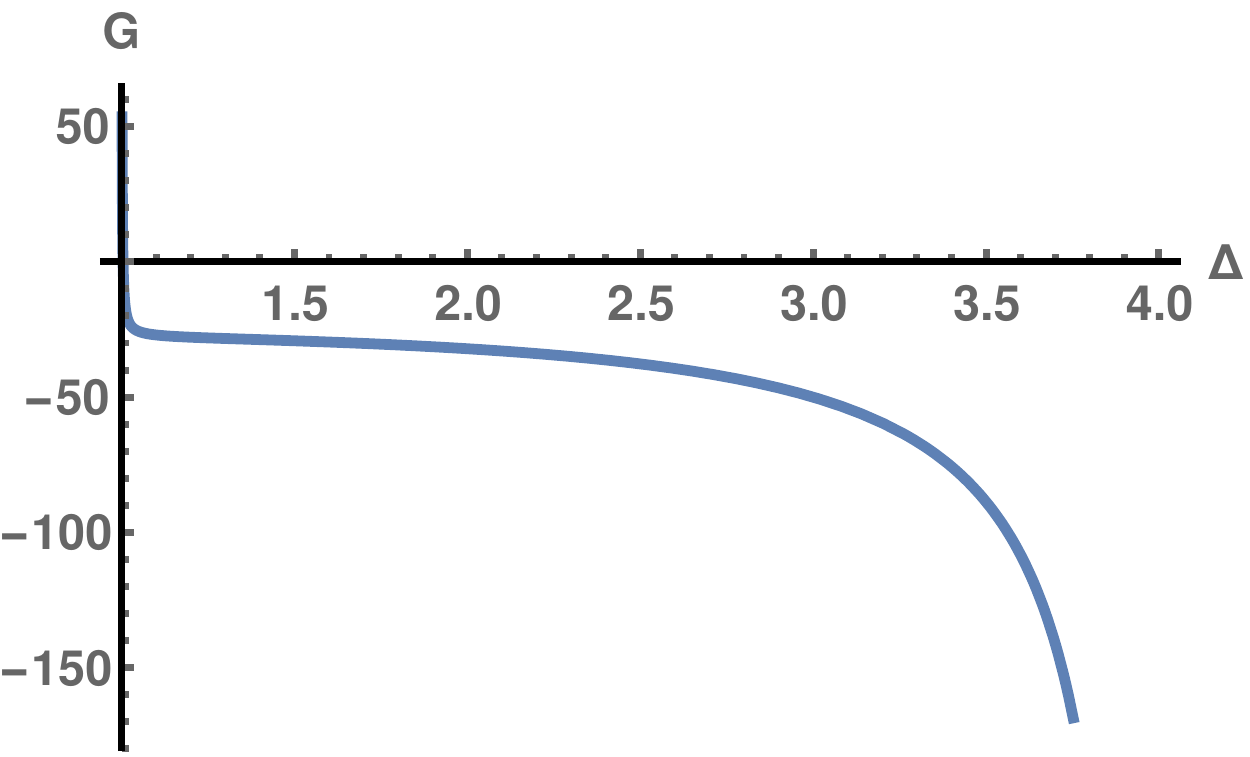}
        }%
        \subfigure[$l=0, ~\Delta>4$ (irrelevant scalar)]{%
           \label{fig:l=0irrel}
           \includegraphics[width=0.4\textwidth]{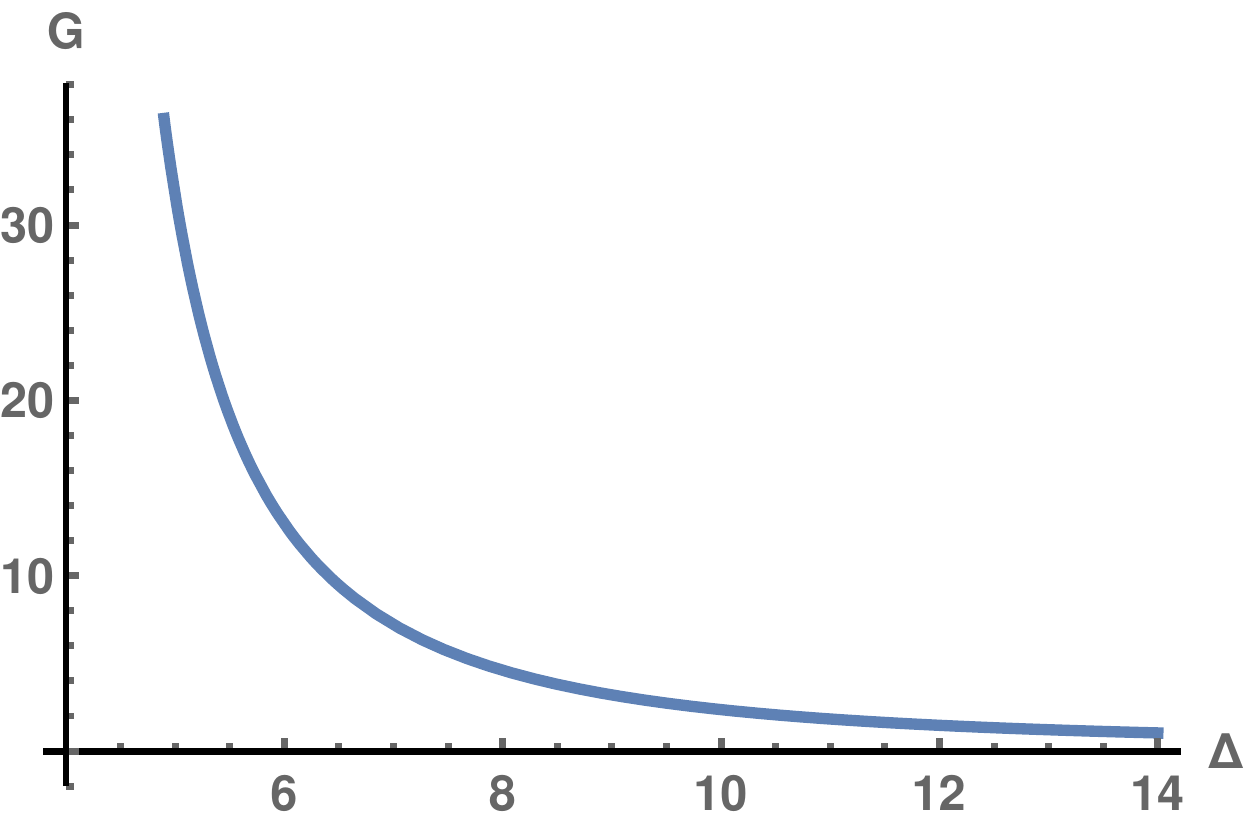}
        }\\ 
        \subfigure[$l=2$]{%
            \label{fig:l=2}
            \includegraphics[width=0.4\textwidth]{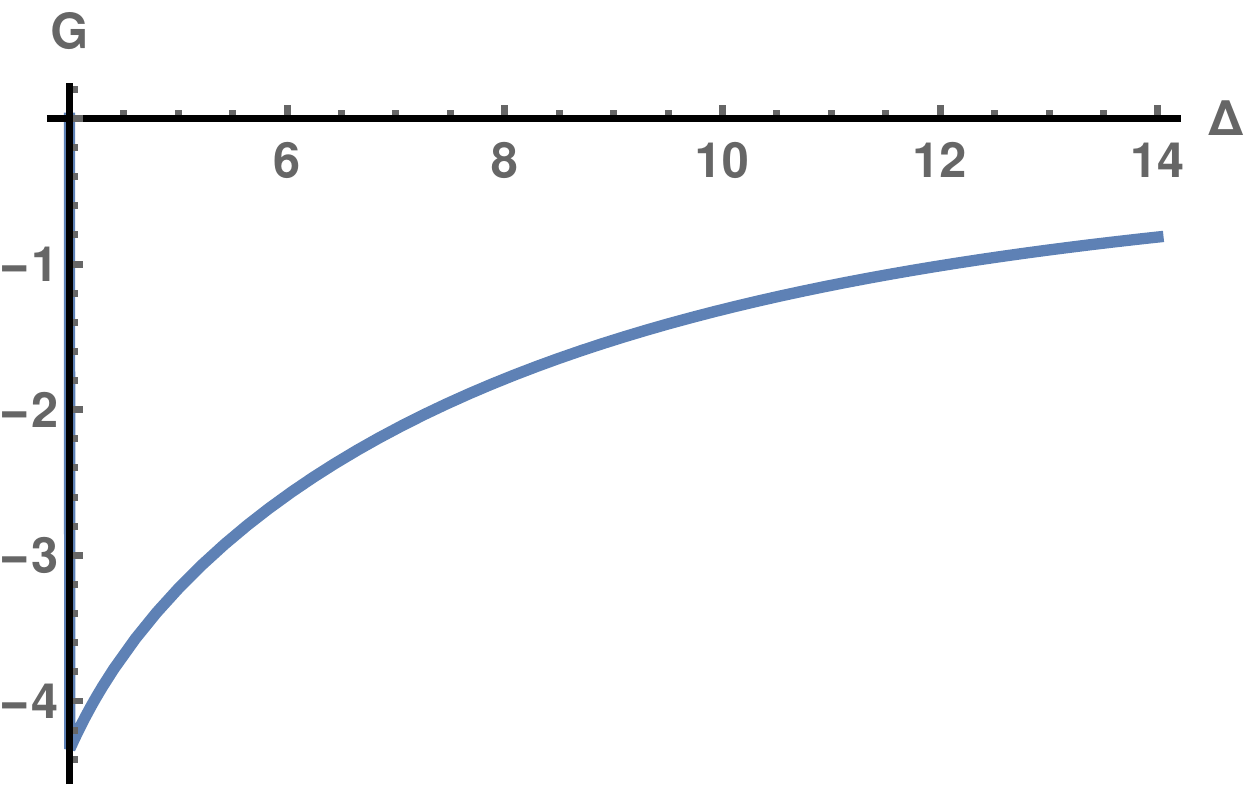}
        }%
        \subfigure[$l=4$]{%
            \label{fig:l=4}
            \includegraphics[width=0.4\textwidth]{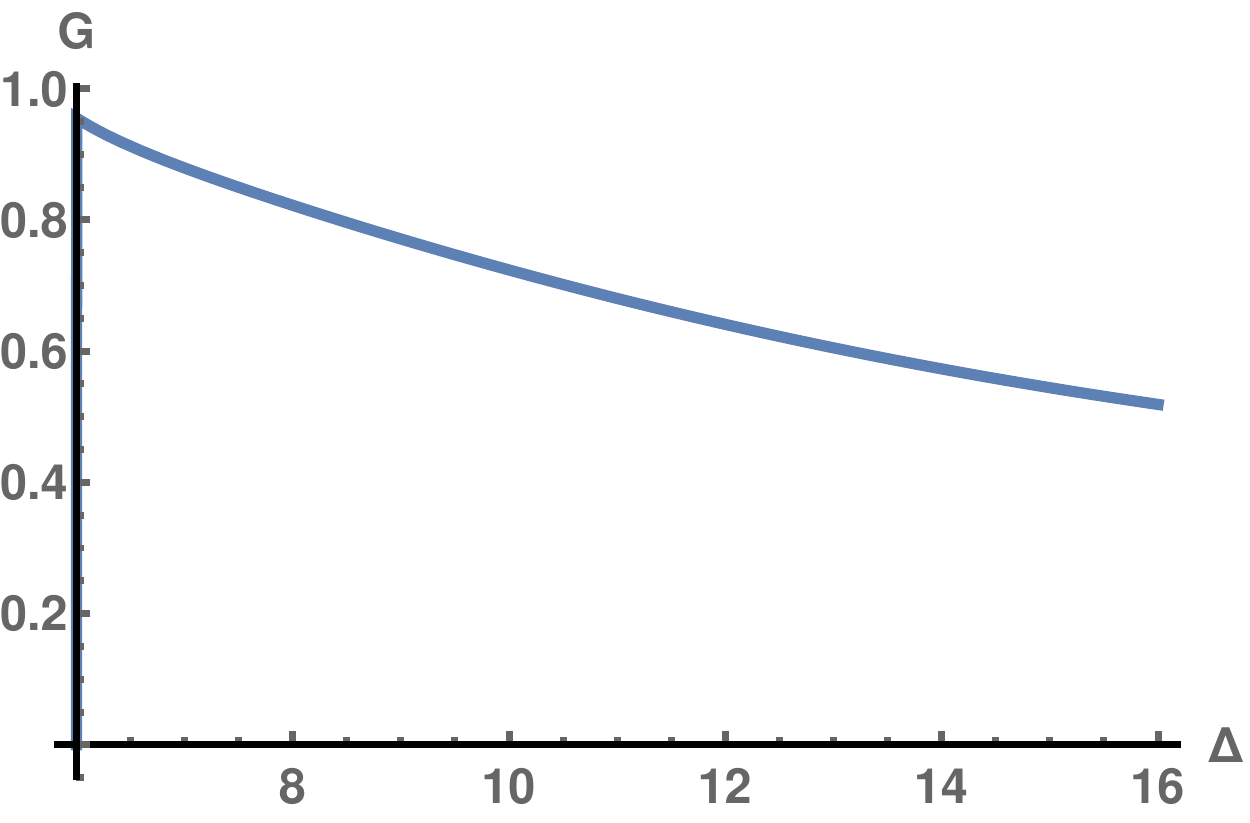}
        }\\ 
        \subfigure[$l=6$]{%
            \label{fig:l=2}
            \includegraphics[width=0.4\textwidth]{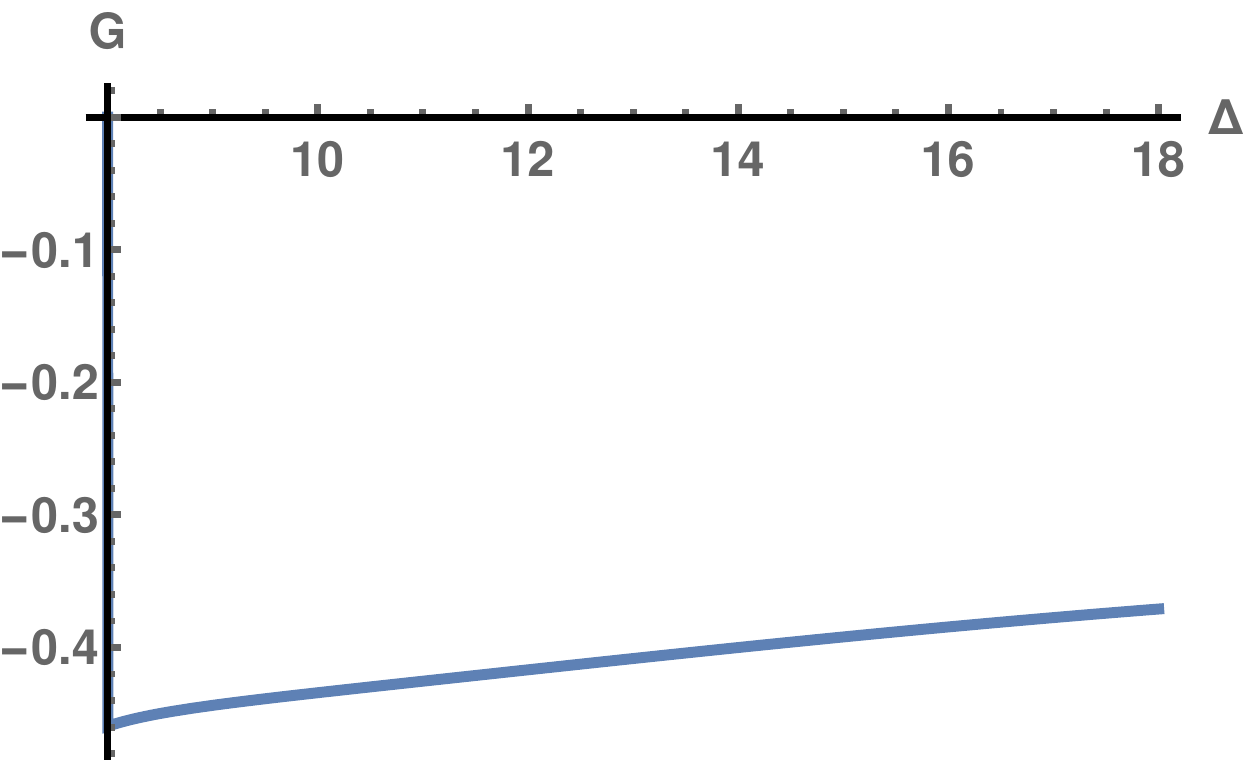}
        }%
        \subfigure[$l=8$]{%
            \label{fig:l=4}
            \includegraphics[width=0.4\textwidth]{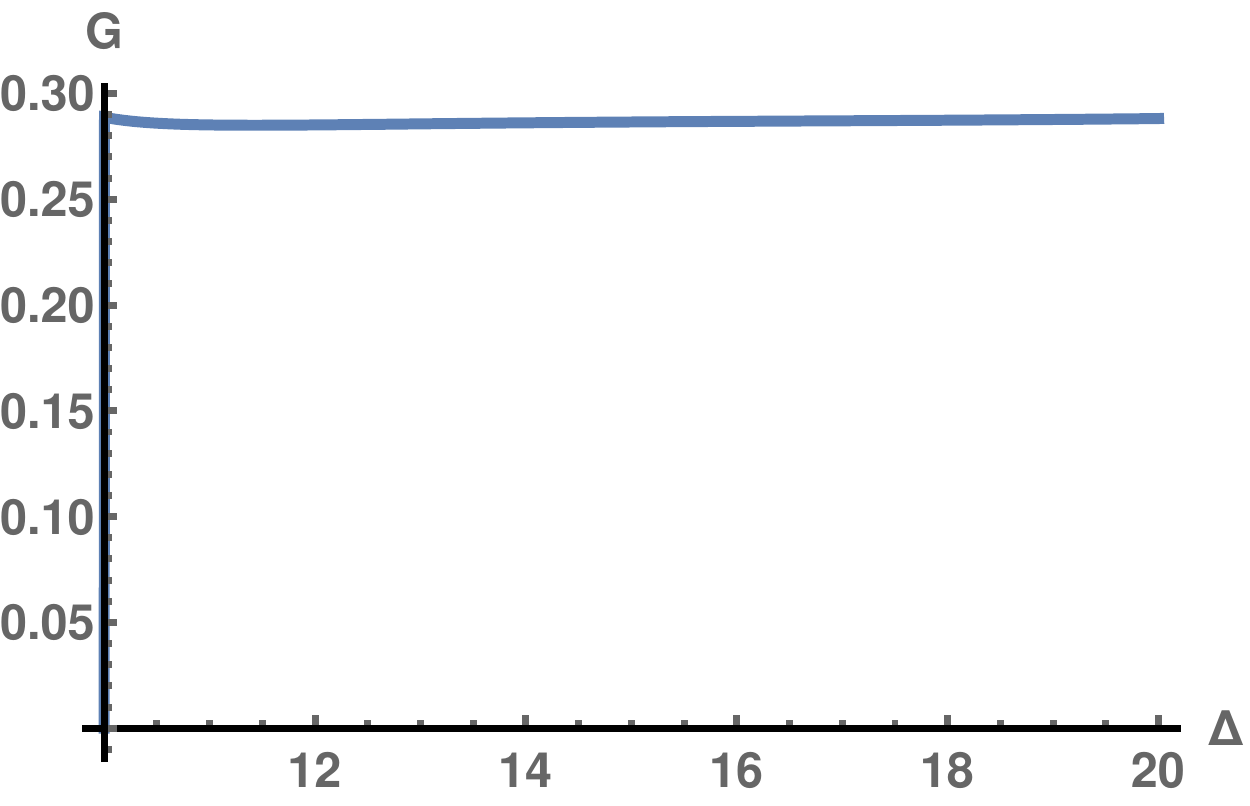}
        }%
    \end{center}
    \caption{%
        Integrated conformal blocks $G$ as a function of operator dimensions for $l=0, ~2, ~4, ~6, ~8$ spin in $d=4$ dimensions.
     }%
   \label{fig:d=2}
\end{figure}

Equation \eqref{sumrule1} is valid in $d$ dimensions, and can be evaluated using known expressions for conformal blocks. Focusing, again, on $d=4$, they read
\be
g_{\Delta,\, l}(z,\bar{z})=\frac{z\bar{z}}{z-\bar{z}}\(K_{\Delta+l}(z)K_{\Delta-l-2}(\bar{z})-K_{\Delta+l}(\bar{z})K_{\Delta-l-2}(z) \),
\ee
where $K_{\beta}$ is given in terms of hypergeometric functions, $K_{\beta}(x)=x^{\beta/2}{}_2F_1\(\tfrac{\beta}{2},\tfrac{\beta}{2},\beta;x\)$. 
From these, one can then compute integrated conformal blocks $G_{\Delta_{\Oc'}, \,l_{\Oc'}}$ defined in eqs. \eqref{intcb1} and \eqref{Icbrel1}. In figure \ref{fig:d=2},  integrated conformal blocks as functions of dimensions $\Delta$ and spin $l$ are provided. Relevant scalar operators have negative integrated conformal blocks and therefore give a negative contribution to the sum rule \eqref{sumrule1}. The opposite holds for irrelevant scalar operators which give instead a positive contribution. All other operators display an alternating behavior: contributions are positive for $l=4,\, 8, \,...$ and negative for  $l=2, \,6, \,...$ (our numerics suggests this behavior to hold for arbitrary values of $l$). One can repeat the above analysis in spacetime dimensions other than four, and it turns out that exactly the same pattern holds. 

A point worth stressing is that the sum rule \eqref{sumrule1} is not unique. For one thing, it depends upon the choice of the integration domain $D$. More generally, this ambiguity comes from crossing symmetry. Indeed, the crossing symmetry equation for a marginal operator is given by
\begin{equation}
\sum_{\Oc'}C_{\cal O O O'}^2\left(v^d ~g_{\Delta_{\Oc'}, \,l_{\Oc'}}(u, v)-u^d ~g_{\Delta_{\Oc'}, \,l_{\Oc'}}(v,u)\right)=0~,
\end{equation}
where $u$ and $v$ are conformal cross-ratios which, in our case, are $u= z \bar z$ and $v=(1-z)(1- \bar z)$. 
For any point $z, \bar{z}$ this gives a sum of the same form as eq.~\eqref{sumrule1} but with a zero on the r.h.s. . Any such sum, or linear combinations thereof, can be added to eq.~\eqref{sumrule1}, modifying the coefficients in front of $C_{\cal O O O'}$'s without changing the r.h.s., hence giving, eventually, a different sum rule. It would be interesting to see whether there exists a choice which makes all terms in the l.h.s. of \eqref{sumrule1} being positive definite. From such a sum rule it would be possible to get very stringent constraints on CFT data as, {\it e.g.}, a lower bound on the central charge of the theory. We were not able to find such linear combination for arbitrary $d$, if it exists at all.

For the sake of what we will do in later sections, let us finally notice that if there are no relevant scalar operators in the ${\cal O O}$ OPE, eq.~\eqref{2lc} simplifies to 
\be
\label{4ps1}
\int d^dx  \langle {\cal O}(0)  {\cal O}(x)  {\cal O}(e)  {\cal O}(\infty)\rangle_c = 0~,
\ee 
and integrated conformal blocks in eq.~\eqref{Icbrel1}, hence contributions as in figure \ref{fig:l=0rel}, would not contribute to \eqref{sumrule1}. Still, this would not change the alternate sign behavior of the sum rule \eqref{sumrule1}, since also operators with $l=2 ~\mbox{mod}~4$ contribute with a negative sign. 

\subsection{Constraints and bounds on CFT data}
\label{3l}

The alternating sign behavior in the sum \eqref{sumrule1} makes it impossible to get straight bounds on $C_{\cal O O O'}$  coefficients, as one might have hoped. Nevertheless, one can still extract useful information out of  \eqref{sumrule1} , as we are going to discuss below.

\subparagraph*{Nearly marginal operators.} Along a conformal manifold the dimension of a generic (that is, non-protected) operator changes continuously as a function of the couplings $g$ parametrizing the conformal manifold. In particular, it can happen that an operator ${\cal K}$ is relevant for $g < g_*$, irrelevant for $g>g_*$, and becomes marginal at $g=g_*$. From \eqref{1lc} it follows that, at $g=g_*$, $C_{\cal O O K}=0$.  
On the other hand, figures \ref{fig:l=0rel} and \ref{fig:l=0irrel} show that integrated conformal blocks of scalar operators blow up when $\Delta \to d$. More precisely, one can see that  
\begin{equation}
G_{\Delta , 0}\propto\frac{1}{\Delta-d} ~~\mbox{when}~~ \Delta \rightarrow d~.
\end{equation}
In order to keep the two-loop beta function coefficient finite, it should be that\footnote{This holds unless for (the very fine-tuned) situations in which there exists a second  marginal operator at $g=g_*$ that changes its dimension from being irrelevant to be relevant, in such a way that the two singularities compensate each other.}
\be\label{behaviour}
\lim_{\Delta \to d}C^2_{\cal O O {\cal K}} \, G_{\Delta, 0} =\;\mbox{finite}~,
\ee
implying that as $\Delta \to d$, $C_{\cal O O {\cal K}}$ must approach zero at least as fast as $(\Delta -d)^{1/2}$. 
 This gives a prediction on how the OPE coefficient approaches zero as a function of $g - g_*$ (in fact, a lower bound on such a dependence). 

The simplest testing ground one can think of to put this prediction at work is ${\cal N}=4$ SYM, which admits an exactly marginal deformation associated to the gauge coupling itself. Indeed, the free theory is part of the conformal manifold and one can work at arbitrary small coupling, where computations can be reliably done. As an example, one can consider the $D$-component (in ${\cal N}=1$ language) of the Konishi multiplet, $\mathcal{K}=\mbox{Tr}\,X^i X_i$, which is marginal at $g_{YM}=0$ and becomes marginally irrelevant in the interacting theory. Its anomalous dimensions is known \cite{Arutyunov:2000im,Bianchi:2001} and one could then give a prediction, via \eqref{behaviour}, on the behaviour of $C_{\Oc \Oc \mathcal{K}}$, where $\Oc$ is the marginal operator dual to the (complexified) gauge coupling. However, due to a $U(1)$ bonus symmetry enjoyed by ${\cal N}=4$ SYM and a corresponding selection rule  \cite{Intriligator}, such OPE coefficient is predicted to vanish. Hence, in this specific case, the constraint \eqref{behaviour}  does not provide any new information.

In fact, ${\cal N}=4$ SYM admits a larger conformal manifold, along which the predictions coming from eq.~\eqref{behaviour} become relevant. Using again an $\cn=1$ notation, ${\cal N}=4$ has three chiral superfields $\F_i$ that transform in the fundamental representation of the $SU(3)$ flavor symmetry. These chiral superfields can be used to construct an exactly marginal $SU(3)$ invariant superpotential (the ${\cal N}=4$ cubic superpotential) and ten classically marginal superpotential terms that transform as a {\bf 10} of $SU(3)$. Two out of the ten marginal superpotentials are exactly marginal  \cite{Leigh:1995ep}. Deforming the $\cn=4$ theory by these exactly marginal operators explicitly breaks the $SU(3)$ symmetry and lifts the dimension of other classically marginal operators along with the $SU(3)$ broken currents. These operators acquire anomalous dimension at the quadratic order in the deformation and were explicitly obtained in \cite{Bashmakov:2016uqk}. The  constraint \eqref{behaviour} then predicts that the OPE coefficient $C_{\cal O O K}$, ($\cal O$ being the $SU(3)$ breaking exactly marginal operator and ${\cal K}$ any of the marginally irrelevant operators) scales at least linearly in the exactly marginal coupling. Note that these statements are independent from $g_{YM}$ so they hold also at strong coupling. Very similar behavior occurs for a large class of ${\cal N}=1$ superconformal quiver gauge theories obtained by considering D-branes at Calabi-Yau singularities \cite{Kol:2002zt,Benvenuti:2005wi,Green:2010da,Kol:2010ub}. There again, non-trivial conformal manifolds exist, along which operators which are marginal in the undeformed theory acquire an anomalous dimensions, which can be computed using similar techniques as for ${\cal N}=4$ SYM (see \cite{Bashmakov:2016uqk} for details). What is interesting in these models is that, unlike ${\cal N}=4$ SYM, there is no point whatsoever on the conformal manifold in which the theory is weakly coupled. So these results are intrinsically at strong coupling.

\subparagraph*{Estimating the tail.} The fact that $A$ in eq.~\eqref{sumrule1} is a positive number implies that the $\cal O O$ OPE must contain at least one operator with positive integrated conformal block. From the results reported in figure \ref{fig:d=2} it follows that at least an irrelevant scalar operator or else a spinning operator with $l=4$ mod 4 must be present. In principle, this can be interesting since to date numerical bootstrap results are less powerful as far as OPE of operators of dimension $\Delta\gtrsim d$ are concerned. When a marginal operator $\Oc$ exists, instead, one gets constraints also about the spectrum of other such operators. This can be seen as follows. 

Let us consider a given value $\Delta=\Delta_*$ and divide the sum \eqref{sumrule1} as 
\begin{equation}
\label{sumsplit1}
\sum_{\Oc': \Delta<\Delta_*} ~C^2_{\cal O O O'} ~G_{\Delta_{\Oc'}, \,l_{\Oc'}} + \sum_{\Oc': \Delta>\Delta_*} ~C^2_{\cal O O O'} ~G_{\Delta_{\Oc'}, \,l_{\Oc'}} = A~.
\end{equation}
Since the series is expected to converge, there should exist (large enough) values of $\Delta_*$ for which
\begin{equation}
\label{delta*1}
\sum_{\Oc': \Delta>\Delta_*} ~C^2_{\cal O O O'} ~G_{\Delta_{\Oc'}, \,l_{\Oc'}} <  A~.
\end{equation}
This means that 
\be
\label{delta*2}
\sum_{\Oc': \Delta<\Delta_*} ~C^2_{\cal O O O'} ~G_{\Delta_{\Oc'}, \,l_{\Oc'}}  > 0~,
\ee
which implies, in turn, that among the operators with dimension $\Delta<\Delta_*$, at least one operator with positive integrated conformal block should exist. If $\Delta_*$ is parametrically large this is something not very informative. If $\Delta_*$ is not too large, instead, one can get interesting constraints on the spectrum of low dimension operators.

\begin{figure}[ht]
\begin{center}
\includegraphics[height=0.25\textheight]{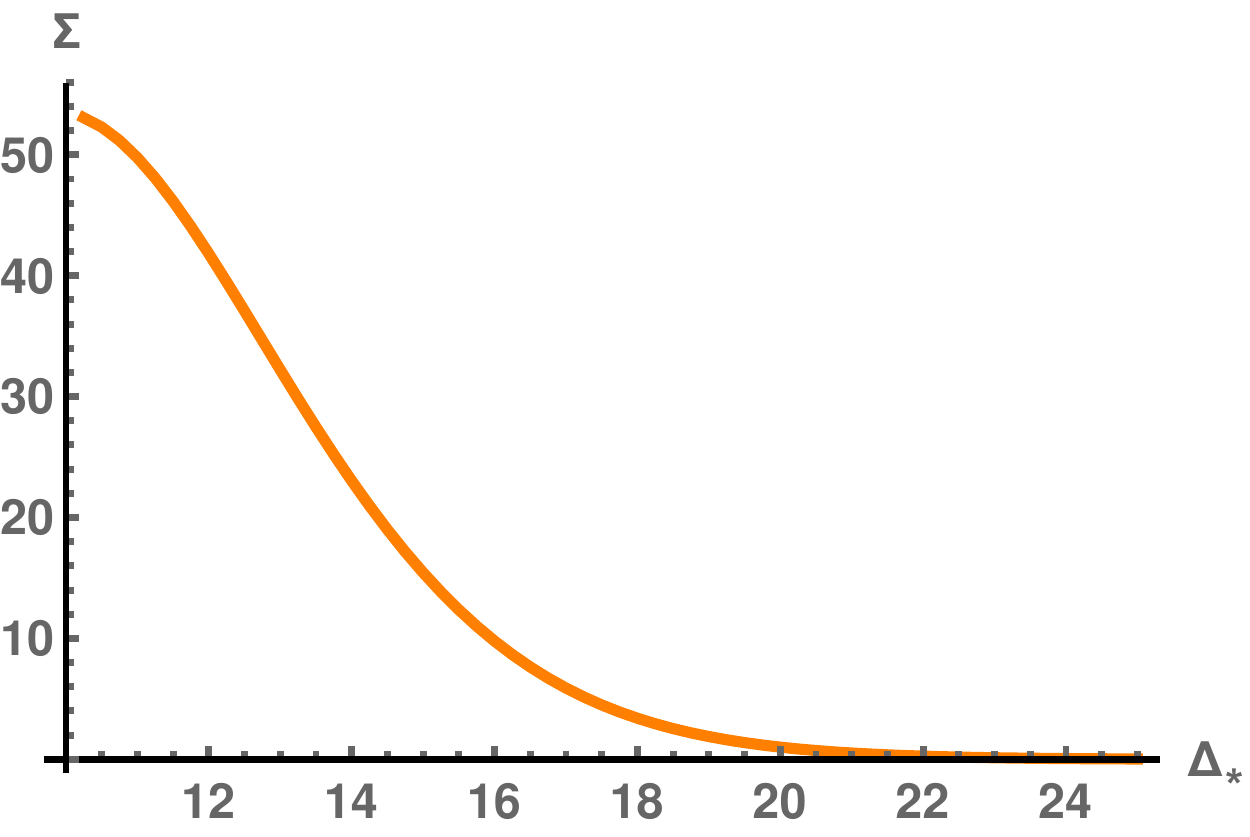}
\caption{The estimate $\Sigma(\Delta_*)$ as a function of $\Delta_*$. }
\label{fig:d=4error}
\end{center}
\end{figure}

One can try to give an estimate of the values of $\Delta= \Delta_*$ for which \eqref{delta*1} is satisfied, {\it e.g.}, using the approach of \cite{Pappadopulo:2012,Rychkov:2015}, where the question of convergence of OPE expansion was addressed, and an estimate of the tail was given. For example, for $d=4$ this takes the form

\begin{equation}
\label{est1}
\sum_{\Oc': \Delta>\Delta_*} ~C^2_{\cal O O O'} g_{\Delta_{\Oc'},l_{\Oc'}}(z, \bar{z})\lesssim \frac{2^{16}\Delta_*^{16}}{\Gamma(17)}\left|\frac{z}{(1+\sqrt{1-z})^2}\right|^{\Delta_*}.
\end{equation} 
One can then define
\begin{equation}
\label{est2}
\Sigma(\Delta_*)\equiv \pi\int_{D} d^2z ~\(\frac{z-\bar{z}}{2i}\)^2\frac{2^{16}\Delta_*^{16}}{\Gamma(17)|z|^8}\left|\frac{z}{(1+\sqrt{1-z})^2}\right|^{\Delta_*}~,
\end{equation}
which means that
\begin{equation}
\label{est3}
 \sum_{\Oc': \Delta>\Delta_*} ~C^2_{\cal O O O'} ~G_{\Delta_{\Oc'}, \,l_{\Oc'}}\lesssim \Sigma(\Delta_*)~.
\end{equation}
The function $\Sigma(\Delta_*)$ is shown in figure \ref{fig:d=4error}.  
In principle, the estimate \eqref{est1} is valid only asymptotically, namely in the limit $\Delta_* \to \infty$. Moreover, the actual value above which the error one is making can be neglected is theory-dependent. Therefore, one should be careful using \eqref{est1} for too low values of $\Delta_*$ and/or to make generic predictions. In fact, numerical bootstrap results suggest that a value of, say, ${\cal O}(10)$, can already be in a safe region for a large class of CFTs (see \cite{Echeverri:2016ztu} for a discussion on this point). 

Looking at  \eqref{delta*2}, it is clear that the lower $\Delta_*$ the more stringent the constraints on low dimension operators. Requiring the l.h.s. of eq.~\eqref{delta*1} to saturate the inequality, which is the best one can do, and evaluate it using \eqref{est3}, we get that $\Sigma(\Delta_*)=A$ for $\Delta_*=16.3$. This is already a large enough value for which the estimate \eqref{est1} can be trusted, for a large class of CFTs \cite{Echeverri:2016ztu}. Looking at figure \ref{fig:d=2} we then conclude that in the OPE of an exactly marginal scalar operator there must be either an irrelevant scalar operator and/or some spin $l=4,8,12$ operators with dimensions  $\Delta \lesssim 16$ (recall that the unitarity bound is $\Delta= d -2 +l $). 

In all above discussion we have been focusing, for definiteness, on $d=4$ dimensions, but similar conclusions can be drawn in any dimensions $d$.

Let us finally note, in passing, that the same approach used here could more generally be used to constrain the spectra of a CFT whenever the two loop $\b$ function coefficient is known.

\section{Conformal manifolds and holography}
\label{holcm}
In this section we want to focus our attention on CFTs admitting a gravity dual description. These can be characterized  as CFTs which admit a large-$N$ expansion and whose single-trace operators with spin greater than two have a parametrically large dimension \cite{Heemskerk:2009pn}. More precisely, in the large-$N$ limit the CFT reduces to a subset of operators  having small dimension ({\it i.e.},  a dimension $\Delta$ that does not scale with $N$), and whose connected $n$-point functions are suppressed by powers of $1/N$. This implies, in particular, that for $N \rightarrow \infty$  the four-point function factorizes and hence the connected four-point function vanishes, like for free operators. However, unlike the latter, these operators, also known as generalized free fields, do not saturate the unitarity bound (see \cite{ElShowk:2011ag} for a nice review). 

Scalar operators are dual to scalar fields in the bulk. From the mass/dimension relation, which (for scalars and in units of the AdS radius) reads 
\be
m^2 = \Delta (\Delta - d)~,
\ee
it follows that in order for the dual operator ${\cal O}$ to be marginal, one needs to consider a massless scalar in the bulk. Its non-normalizable mode acts as a source for ${\cal O}$, and thus corresponds to a deformation in the dual field theory described by eq.~\eqref{conman1} (in other words, the non-normalizable mode is dual to the coupling $g$). The conformal manifold ${\cal M}_c$ is hence mapped into the moduli space  ${\cal M}$ of AdS vacua of the dual gravitational theory, {\it i.e.}, AdS solutions of bulk equations of motion parametrized by massless, constant scalar fields \cite{Tachikawa:2005tq}. 

The duality between ${\cal M}_c$ and ${\cal M}$ makes it manifest the difficulty to have conformal manifolds in absence of supersymmetry. A non-supersymmetric CFT is dual to a non-supersymmetric gravitational theory. Differently from supersymmetric moduli spaces, non-supersymmetric moduli spaces are expected to be lifted at the quantum level. Quantum corrections in the bulk are weighted by powers of $1/N$. Hence, one would expect that a moduli space of AdS vacua existing at the classical level, would be lifted at finite $N$. 

For theories with a gravity dual description, this is the simplest argument one can use to argue that conformal manifolds without supersymmetry are something difficult to achieve. In this respect, it is already interesting to find non-supersymmetric conformal manifolds persisting at first non-planar level. One of our aims, in what follows, is to show that this is not an empty set. 

We will consider the simplest model one can think of, namely a massless scalar field $\f$ minimally coupled to gravity. This  corresponds to CFTs which, as far as single-trace operators are concerned, in the large-$N$ limit reduce to a single low-dimension scalar operator ${\cal O}$, dual to $\f$.\footnote{A CFT must include the energy-momentum tensor. Our toy-model could be thought of as a sector of an AdS compactification in which there is a self-interacting scalar in the approximation that gravity decouples, as in {\it e.g.} \cite{Heemskerk:2009pn}. Most of what we will do, does not depend on this approximation.}

\subsection{Conformal perturbation theory and the 1/N expansion}

Our first goal is to discuss how the two perturbative expansions we have to deal with in the CFT, that is, conformal perturbation theory, which is an expansion in $g$, and the $1/N$ expansion, are related to one another from a holographic dual perspective.

Let us consider a bulk massless scalar $\f$ having polynomial interactions of the form
\be
\label{scalintgen1}
\sum_{n} \l_n \[ \f^n\]~,
\ee
where $n \geq 3$ and $\[ \f^n\]$ stands for Lorentz invariant operators made of $n$ fields $\f$'s. For the time being, we do not need to specify their  explicit form, which can also include derivative couplings. 
  
\begin{figure}[ht]
\begin{center}
\includegraphics[height=0.17\textheight]{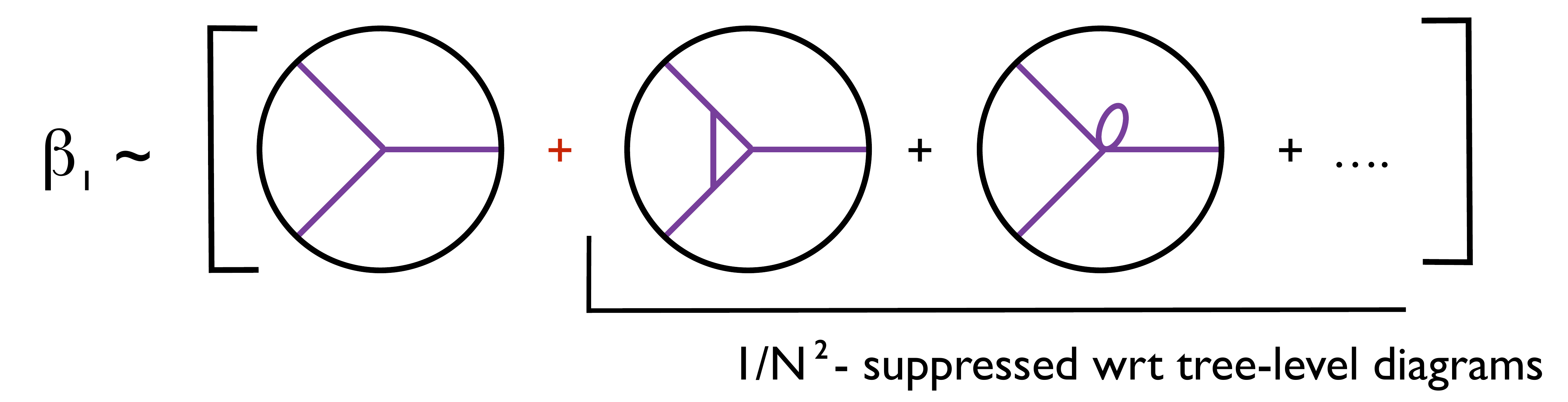}
\caption{Witten diagrams contributing  to $C_{\cal OOO}$. Violet lines correspond to propagation of $\f$ fields and may have spacetime derivatives acting on them, depending on the specific structure of the operators \eqref{scalintgen1}. At tree-level, only cubic couplings can contribute to the three-point function. At loop level, also couplings with $n>3$ can contribute, {\it e.g.}, the quintic coupling shown in the figure.}
\label{Wdt3all}
\end{center}
\end{figure}

Let us consider the one-loop coefficient $\beta_1$, eq.~\eqref{1lg}. In order to compute it holographically, one needs to evaluate Witten diagrams \cite{Witten:1998qj} with three external lines. Witten diagrams are weighted with different powers of $1/N$, corresponding to tree-level and loop contributions in the bulk. As shown in figure  \ref{Wdt3all}, at tree level only the cubic vertex can contribute to the three-point function. At higher loops, instead, also couplings with $n>3$ may contribute to $\b_1$.  

\begin{figure}[ht]
\begin{center}
\includegraphics[height=0.21\textheight]{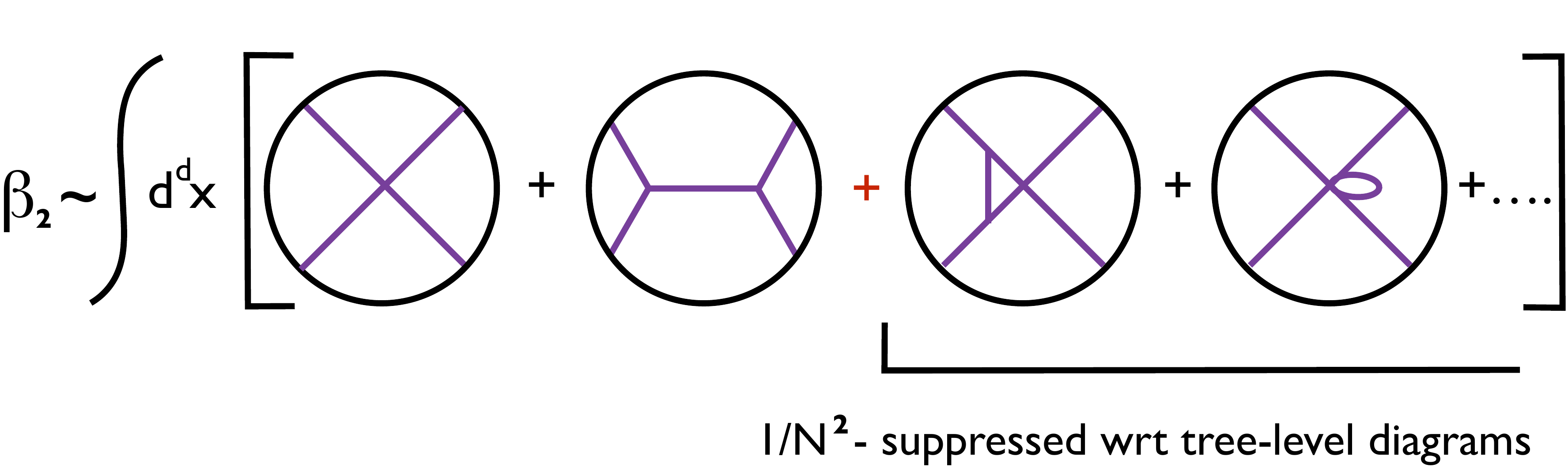}
\caption{Structure of Witten diagrams contributing to the two-loop coefficient of $\beta(g)$, after integration in $d^dx$. Conventions are as in figure \ref{Wdt3all}.}
\label{Wdt4all}
\end{center}
\end{figure}

A similar story holds for the two-loop coefficient $\beta_2$ (note that in our one-field model eq.~\eqref{2lc} simplifies just to the integral of the four-point function, eq.~\eqref{4ps1}). To leading order, there are two contributions. The contact quartic interaction and the cubic scalar exchange, as shown in figure  \ref{Wdt4all}. Again, at higher-loops in the bulk coupling, one can get contributions also from operators with $n >4$. 

The analysis applies unchanged to the three-loop coefficient $\beta_3$ and higher. In particular,  only operators $\[ \f^n\]$ with $n \leq m$ can contribute to the $m$-point function of ${\cal O}$ at tree level. Conversely, at loop level, also operators with $n>m$ may contribute. 

What we would like to emphasize with this discussion is that by doing tree-level computations in the bulk, one can extract the leading, planar contribution to $\beta(g)$ at all loops in $g$. In other words, classical gravity provides an exact answer, in conformal perturbation theory, to the existence of a conformal manifold, at leading order $1/N$. To get this, rather than computing Witten diagrams, it is clearly much simpler to solve bulk equations of motion and see which constraints on the structure of the operators \eqref{scalintgen1} does the existence of AdS solutions with constant $\f$ impose. This is what we will do, first. Then, we will compute explicitly tree-level Witten diagrams contributing to $\b_1$ and $\b_2$, and check that the constraints one gets by requiring them to vanish, are in agreement with those coming from equations of motion analysis. 

A non-trivial question one can ask is whether the vanishing of $\beta(g)$ at two-loops leaves some freedom in the scalar couplings compared to the equation of motion analysis. And, if this is the case, at which loop order in CPT one should go, to fix such freedom. The answer turns out to be rather simple: admissible operators of the form $[\f^n]$ will be fully determined by imposing the vanishing of the $\b$-function coefficient $\b_{n-2}$, and no higher orders will be needed. The toy model we are going to discuss has operators with $n=3,4$ only, and, consistently, we will see that the constraints coming just from the vanishing of $\b_1$ and $\b_2$, will provide the full gravity answer.

Another interesting question is which further constraints the vanishing of the one and two-loop coefficients of $\beta(g)$ put on the CFT taking into account $1/N$ corrections, that is, going beyond planar level. As already emphasized, one does not expect exact conformal manifolds to survive at finite $N$, without supersymmetry. However, one can ask whether non-trivial CFTs with non-supersymmetric conformal manifolds persisting at first non-planar level could exist. That this can be, it is not obvious, and this is what we will address next.

\subsection{Scalar fields in AdS}

We want to compare the holographic analysis with CPT at two-loops, which, as such, involves at most four-point functions, eqs. \eqref{1lc} and \eqref{4ps1}. Therefore, for simplicity, we will focus on models with cubic and quartic couplings, only. The bulk action reads
\be
\label{bulkgen1}
S = \frac 1{2 \kappa^2_{d+1}} \int d^{d+1}x \sqrt{-g} \left(R - \frac 12 g^{\mu\nu} \partial_\m \phi \,\partial_\n \phi - 2 \Lambda + \[ \phi^3\] + \[ \phi^4\] \right)~,
\ee
where $\Lambda$ is the (negative) cosmological constant and the last two terms represent cubic and quartic interactions. The absence of a mass term for $\phi$ guarantees that the dual operator ${\cal O}$ is marginal, {\it i.e.} $\Delta_{\Oc}=d$. We would like to constrain the explicit form of cubic and quartic couplings by requiring the existence of a conformal manifold under a deformation parametrized by $\phi$ itself.  We take $\k_{d+1}\sim N^{-1}$ to match holographic correlators with CFT correlation functions in the large-$N$ limit. In the above normalization, the two point function $\la \co\co \ra$ scales as $N^2$. Such unusual normalization has the advantage to treat democratically all Witten diagrams (as well as the dual $n$-point functions, and so the $\beta$-function coefficients $\b_n$), in the sense that, regardless the number of external legs, they all scale the same with $N$, at any fixed order in the bulk loop expansion.\footnote{The interested reader can explicitly check this statement, after having properly chosen the normalization of the bulk-to-boundary propagator.} This is the most natural choice that avoids mixing-up the expansion in $1/N$ with that in $g$.

From the action \eqref{bulkgen1} one can derive the equations of motion, which read 
\bea
\label{Ein1}
R_{\m\n} - \frac 12 g_{\m\n} R &=& \frac 12 \partial_\m\f \partial_\n \f - \frac 12 g_{\m\n} \,( \frac 12 \partial_\rho\f \,\partial^\rho \f  + 2 \Lambda) - 
\frac{1}{\sqrt{-g}} \frac{\delta}{\delta g_{\m\n}} \sqrt{-g}  \left( \[ \phi^3\] + \[ \phi^4\] \right)
\ \nn \\
\\
\label{Scal1}
\Box_g \, \phi &=&  - \frac{\delta}{\delta \f} \left(\[ \phi^3\] + \[ \phi^4\]\right)~,
\eea
where $\Box_g = g^{\m\n} \nabla_\m \nabla_\n = g^{\m\n} (\partial_\m \partial_\n - \Gamma^\r_{\m\n} \partial_\r)$.

We need to look for pure AdS solutions with constant scalar profile. In absence of interactions, that is in the strict generalized free-field limit, the large-$N$ CFT reduces to a massless free scalar $\phi$ propagating in a rigid AdS background. The equations of motion admit a solution with AdS metric and constant scalar field $\f = \f_0$ which, in Poincar\'e coordinates, reads
\bea
\label{AdSmetric}
ds^2 &=& \frac{L^2}{z^2} \left(dz^2 + dx_i dx^i \right) \\
\f &=& \f_0
\eea 
with $L=\sqrt{d (1-d) \Lambda}$ being the AdS radius and the AdS boundary sitting at $z=0$. The modulus $\f_0$ parametrizes the dual conformal manifold, described by the deformation $g \int d^dx \,{\cal O}$. Eqs.~~\eqref{1lc} and \eqref{4ps1} are trivially satisfied: since $\f$ is a free field, Witten diagrams vanish identically (in particular, in eq.~\eqref{4ps1} the integrand itself vanishes). 

Let us now consider possible cubic and quartic interactions. From eqs.~\eqref{Ein1}-\eqref{Scal1} it follows that couplings compatible with solutions with AdS metric and a constant scalar profile are couplings where spacetime derivatives appear (note that, due to Lorentz invariance, only even numbers of derivatives are allowed). Schematically, acceptable operators look like
\be
\label{genop1}
\nabla \nabla \dots \f \; \nabla \nabla \dots \f \; \nabla \nabla \dots \f \nabla \nabla \dots \f  \dots ~,
\ee
where full contraction on Lorentz indexes is understood and some (but not all) naked $\f$'s, that is $\f$'s without derivatives acting on them, can appear. Therefore, at the classical level, {\it i.e.} to leading order in $1/N$, the requirement of existence of a conformal manifold under the deformation \eqref{conman1} rules out the non-derivative couplings $\f^3$ and $\f^4$, only.\footnote{One can consider the more general structure \eqref{scalintgen1} and the same conclusion holds. Any coupling $[\f^n]$ with (an even number of) derivatives is allowed, classically.} 

As anticipated, we want to compare the above analysis with a direct computation of three and (integrated) four-point functions, which are related to the one and two-loop coefficients of $\b(g)$ via eqs.~\eqref{1lg}-\eqref{2lg}, by means of tree-level Witten diagrams. This could be seen as a simple AdS/CFT  self-consistency check, but one can in fact learn from it some interesting lessons, which could be useful when considering more involved models, as well as when taking into account loop corrections in the bulk.

\subsubsection{Tree-level Witten diagrams}
\label{treelevelsec}

\begin{figure}[ht]
\begin{center}
\includegraphics[height=0.14\textheight]{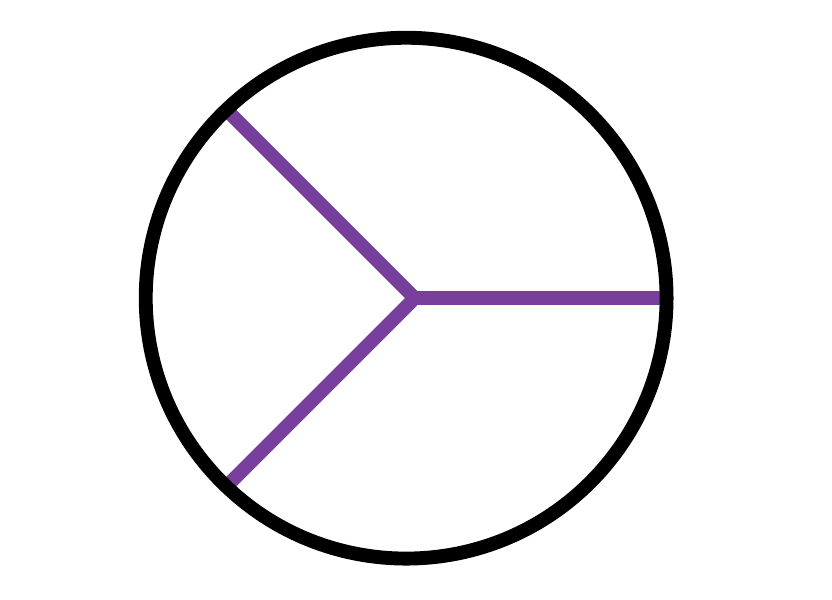}
\caption{Witten diagram contributing  to $\b_1$  at leading order in $1/N$.}
\label{Wdt3}
\end{center}
\end{figure}

Let us consider the one-loop coefficient $\b_1$, which is proportional to $C_{\cal OOO}$. To leading order at large $N$, this corresponds to the Witten diagram shown in figure \ref{Wdt3}, to which only cubic couplings $[\f^3]$ can contribute.

The pure non-derivative coupling $\f^3$ provides a non-vanishing contribution to $C_{\cal OOO}$. Therefore, it is excluded. The first non-trivial couplings are then two-derivative interactions. In principle, the following interaction terms are allowed
\be
\label{2dercp}
\f \nabla_\m \f \nabla^\m \f~~,~\f^2  \, \nabla_\m \nabla^\m \f~.
\ee
Upon using integration by parts and the equation of motion which, at lowest order in the couplings, is just $\nabla_\m\nabla^\m \f= 0$, these interactions are either total derivatives or vanish on-shell. Therefore, they do not contribute to $C_{\cal OOO}$ (this is to be contrasted with the case of a massive scalar, where these interactions are proportional to $\phi^3$). 

Next, one can consider interactions with four spacetime derivatives, that is
\be
\label{4dercp0}
\f \nabla_\m \nabla_\n \f \nabla^\m \nabla^\n \f~~,~~\nabla_\m \f  \nabla_\n \f \nabla^\m \nabla^\n \f~~,~~\f^2 \nabla_\m \nabla_\n \nabla^\m \nabla^\n \f~.
\ee
These terms are also either vanishing on-shell or total derivatives, and do not provide any contribution to the three-point function $\la \cal OOO\ra$, at leading order. Let us briefly see this. Using integration by parts, the second term in \eqref{4dercp0} can be written as
\be
\int ~\de_\m \f \de_\n \f \de^\m \de^\n\f = -\frac12 \int~  \nabla_\n\nabla^\n\f ~ \de_\m \f \de^\m \f~, \\ 
\ee
which vanishes upon using the equation of motion. As for the other two terms in \eqref{4dercp0}, using the identity $[\Box, \nabla_\mu]\phi=-d~\nabla_\mu\phi$, they can be re-written, respectively, as 
\begin{align}
\int~  \f \de_\m \de_\n \f \de^\m \de^\n\f &= \int  \( \frac12 ~ \nabla_\n\nabla^\n\f ~ \de_\m \f \de^\m \f-\frac d2~  \f^2 \nabla_\n\nabla^\n \f\)~,\\
\f^2 \nabla_\m \nabla_\n \nabla^\m \nabla^\n \f &= \f^2 \nabla_\m \nabla^\m \nabla_\n \nabla^\n \f - d\f^2 \nabla_\m \nabla^\m \f~.
\end{align}
Again, both terms vanish upon using the equation of motion, and hence provide no contribution to $C_{\cal OOO}$.  
One can proceed further, and consider couplings  with an increasing number of derivatives, with structures that generalize \eqref{4dercp0}. Using  previous results and proceeding by induction, one can  prove that contributions vanish for any number of derivatives. The upshot is that all operators with two or more derivatives either vanish or can be turned into total spacetime derivatives, and hence give a vanishing contribution to the Witten diagram in figure \ref{Wdt3} and, in turn, to $C_{\cal OOO}$. 

\begin{figure}[ht]
\begin{center}
\includegraphics[height=0.14\textheight]{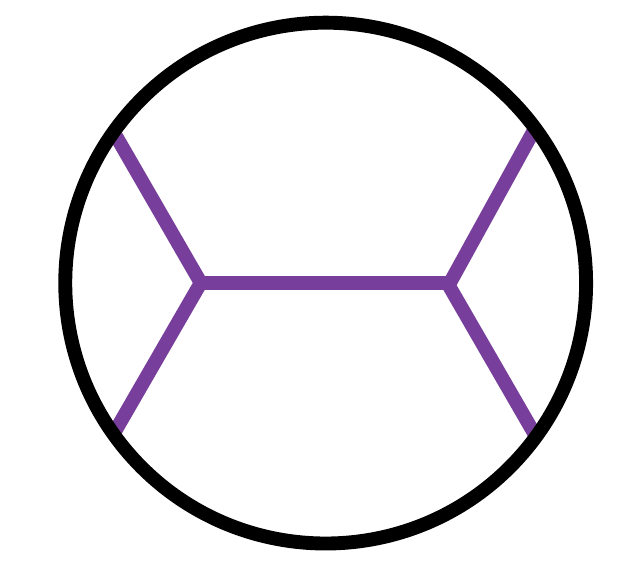}
\caption{Exchange Witten diagram contributing to $\b_2$, after integration in $\int d^dx$.}
\label{Wdt4cubic}
\end{center}
\end{figure}

\vskip 6pt
Although derivative couplings provide a vanishing contribution to cubic Witten diagrams, they can provide non-vanishing contribution to the four-point function by exchange Witten diagrams like the one depicted in figure \ref{Wdt4cubic} (which include, in the dual CFT, the exchange of double-trace operators). Therefore, these interactions can potentially contribute to the two-loop coefficient $\b_2$. 

The pure non-derivative coupling $\f^3$ is already excluded by previous analysis (and it would also contribute to the Witten diagram in figure \ref{Wdt4cubic}, in fact). Let us then start considering contributions from operators having one field $\f$ not being acted by derivatives, {\it i.e.} the first ones in \eqref{2dercp} and \eqref{4dercp0} and generalizations thereof, that is operators of the form
\be
\f \, \nabla \nabla \dots \f \, \nabla \nabla \dots \f ~.
\ee
There are two possible types of exchange Witten diagrams: (a) diagrams where all external lines are acted by derivatives, (b) diagrams where at least one external line is free of derivatives. Focusing, for definiteness, on two-derivative couplings, contributions of type (a) and (b) correspond to the following integrals, respectively
\begin{align}
\int d^d x_1 \int d^d w_1 dz_1  \int d^d w_2 dz_2 &~\nabla_\m K(z_1,w_1-x_1)\nabla^\m K(z_1,w_1-x_2)G(z_1-z_2, w_1- ,w_2)\NO\\
& \nabla_\n K(z_2,w_2-x_3)\nabla^\n K(z_2,w_2-x_4)\label{exchangeA}\\
\int d^d x_1 \int d^d w_1 dz_1  \int d^d w_2 dz_2 &~K(z_1,w_1-x_1)\nabla^\m K(z_1,w_1-x_2)\nabla_\m^{(1)}\nabla_\n^{(2)} G(z_1-z_2, w_1- ,w_2)\NO\\
& K(z_2,w_2-x_3)\nabla^\n K(z_2,w_2-x_4)~.\label{exchangeB}
\end{align}
$K(z,w-x_i)$ is the bulk-to-boundary propagator which, for massless scalars, reads
\be
K(z,w-x)=\(\frac{z}{z^2+(w-x)^2}\)^d~,
\ee
and satisfies the equation $\nabla_{\mu}\nabla^{\mu} K(z,w-x) = \Box_g K(z,w-x) = 0$. $G(z_1-z_2 , w_1-w_2)$ is instead the bulk-to-bulk propagator which, for massless scalars, reads 
\be
G(z_1 - z_2,w_1-w_2)=\frac{2^{-d}C_d}{d}\x^d F\(\frac d2,\frac d2+\frac12;\frac d2+1;\x^2\)~,~~~~C_d=\frac{\Gamma(d)}{\p^{d/2}\Gamma(d/2)}~,
\ee
where $\x$ is the geodesic distance between the two points in the bulk where interactions occur, $(z_1 , w_1)$ and $(z_2 , w_2)$, 
\be
\x=\frac{2 z_1z_2}{z_1^2+z_2^2+(w_1-w_2)^2}~.
\ee
The bulk-to-bulk propagator satisfies the equation $\Box_g G(z_1,w_1;z_2,w_2)=\frac{1}{\sqrt{g}}\d\(z_1-z_2,w_1-w_2\)$. 

Diagrams of type (a) vanish because the integrated bulk-to-boundary propagator $K(z,w-x)$ is independent of $z$ and $w$, namely
\be\label{xIntegral}
\int d^dx ~K(z,w-x)= \frac{\p^{d/2}\G(d/2)}{\G(d)}~,
\ee
and, plugging \eqref{xIntegral} into \eqref{exchangeA}, one gets 
\be
\label{intdK}
\int d^d x_1 \nabla_\m K(z_1,w_1;x_1)=0~.
\ee
Diagrams of type (b), after $x$-integration, also vanish. Indeed, the integral \eqref{exchangeB} becomes
\begin{align}
\label{btype1}
\frac{\p^{d/2}\G(d/2)}{\G(d)}\int d^d w_1 dz_1  \int d^d,w_2 dz_2 &~\nabla^\m K(z_1,w_1-x_2)\nabla_\m^{(1)}\nabla_\n^{(2)} G(z_1-z_2 , w_1-w_2)\NO\\
& K(z_2,w_2-x_3)\nabla^\n K(z_2,w_2-x_4)~,
\end{align}
and, integrating by parts, one can transfer the covariant derivative $\nabla_\m^{(1)}$ acting on the bulk-to-bulk propagator onto $K(z_1,w_1-x_2)$, getting
\begin{align}
\label{btype2}
-\frac{\p^{d/2}\G(d/2)}{\G(d)}\int d^d w_1 dz_1  \int d^d w_2 dz_2 &~\Box_g K(z_1,w_1-x_2)\nabla_\n^{(2)} G(z_1 - z_2 ,w_1-w_2)\NO\\
& K(z_2,w_2-x_3)\nabla^\n K(z_2,w_2-x_4)~,
\end{align}
which vanishes because $\Box_g K=0$. This computation can be repeated for terms with four or more derivatives, just  replacing single derivatives acting on the propagators in \eqref{exchangeA} and \eqref{exchangeB} with multiple derivatives. The end result can again be shown to be zero. 
 
The second possible cubic vertexes which could contribute to the exchange Witten diagram are those with derivatives acting on one field only, schematically
\be
\f^2  \, \nabla \nabla \nabla \dots \f~. 
\ee
Using properties of Ricci and Riemann tensors in AdS, one can show that these couplings can be re-written as sums of terms of the form $\f^2 \,\Box^p \f$, with $p$ an integer. Due to the property $\Box_g K=0$, if derivatives are acting on at least one external line, the result is zero. If not, namely if derivatives act only on the bulk-to-bulk propagator, then the corresponding diagram is a special instance of a (b)-type diagram previously discussed and, following similar steps as in  eqs.~\eqref{btype1}-\eqref{btype2}, one gets again a vanishing result. 

Finally, let us consider shift-symmetric couplings, that is couplings without naked $\f$'s. This kind of couplings give rise to diagrams of type (a), very much like \eqref{exchangeA}, where all external lines (in fact any line) contain derivatives. Therefore, they do not contribute to exchange Witten diagrams, either. 

This ends our analysis of cubic operators, which fully agrees with equations of motion analysis.

Let us emphasize that while all cubic couplings but $\f^3$ do not contribute at the level of three-point functions, they do, in general,  as far as exchange Witten diagrams are concerned. There, it matters that, in computing the two-loop coefficient $\b_2$, integration in $d^dx$ is required, and this plays a crucial role in providing a vanishing result, in the end. 

\vskip 6pt

\begin{figure}[ht]
\begin{center}
\includegraphics[height=0.14\textheight]{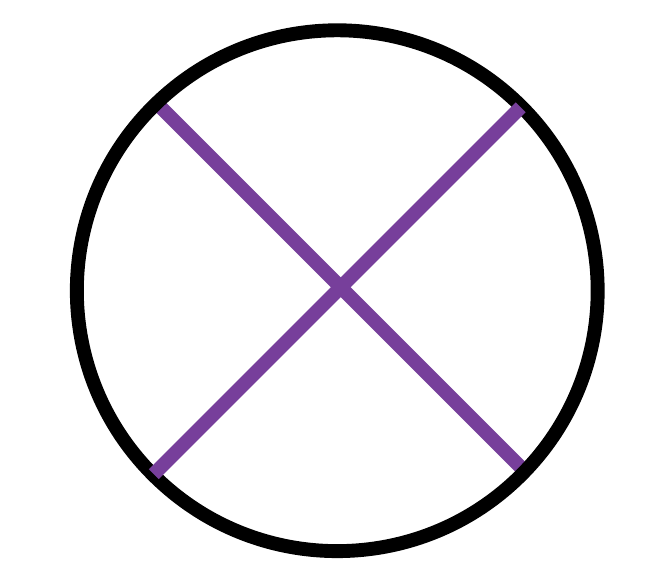}
\caption{Contact Witten diagram contributing  to $\b_2$, after integration in $d^dx$.}
\label{Wdt4quartic}
\end{center}
\end{figure}

Let us now consider quartic couplings. At tree level they do not contribute to $\b_1$, but they can contribute to $\b_2$, instead, via contact-terms, as the one depicted in figure \ref{Wdt4quartic}. 

The operators we should consider are just obtained by adding an extra field $\f$ to all cubic vertexes previously considered. Again, the pure non-derivative coupling $\f^4$ is excluded from the outset, since it clearly gives a non-vanishing contribution. The other operators have the following structures 
\be
\label{quartic1}
\f \, \nabla \nabla \dots \f \, \nabla \nabla \dots \f \, \nabla \nabla \dots \f ~~,~~\f^2 \, \nabla \nabla \dots \f \, \nabla \nabla \dots \f ~~,~~\f^3 \, \nabla \nabla \dots \f ~, 
\ee
as well as the shift-symmetric one 
\be
\label{quartic2}
 \nabla \nabla \dots \f \, \nabla \nabla \dots \f \, \nabla \nabla \dots \f  \, \nabla \nabla \dots \f ~.
\ee
Given our previous analysis it is not difficult to compute the contribution of these diagrams to the integrated four-point function and hence to the $\b$ function two-loop coefficient $\b_2$. Upon integration, the diagram in figure \ref{Wdt4quartic} either gives zero, when the $x$-dependence is on a line where bulk derivatives act, see eq.~\eqref{intdK}, or, after $x$-integration, it reduces to the effective vertex of one of the cubic vertices discussed previously, which vanish. We thus see that all operators \eqref{quartic1} and \eqref{quartic2} do not give any contribution to $\b_2$. Note, again, that $x$-integration plays a crucial role.

To summarize, the constraints on cubic and quartic couplings coming from CPT at two-loops, already capture the (full) gravity answer, as anticipated. From the analysis in section \ref{cpt}, it is not difficult to get convinced that operators with $n$ fields $\f$ will be univocally fixed by computing tree-level Witten diagrams with $n$ external legs, which contribute to the $\b$ function at $n-2$ loop order.  

As already emphasized, a CFT must include the energy-momentum tensor in the spectrum of primary operators, which amounts to include dynamical gravity in the bulk. At tree-level, this would contribute to the exchange Witten diagram in figure \ref{Wdt4cubic}, since now also graviton exchange should be considered in the bulk-to-bulk propagator. For a minimally coupled scalar, which is the case here, the only such contribution would arise from a vertex of the following kind
\be\label{e1}
 h^{\m\n} \pa_\m\f \, \pa_\n\f~,
\ee
where $h_{\m\n}$ denotes the fluctuations of the $AdS_{d+1}$ metric. It is not difficult to see that, because the scalar field $\f$ enters under derivatives, the integrated four-point function is of (a)-type, following our previous terminology, and it vanishes, because of eqs. \eqref{xIntegral} and \eqref{intdK}. 
So, our conclusions are unchanged also once gravity is taken into account.\footnote{For non-minimally coupled scalars one could have other operators contributing to the exchange Witten diagram. Couplings of the type, {\it e.g.}, 
$ R^{\m\n} \pa_\m\f\, \pa_\n\f~$ would again be allowed since the resulting integrated four-point function would also be of the (a)-type. Conversely, couplings like $R \,\f^2$ (and, more generally, any non-derivative coupling) would not be permitted because they would instead contribute to the integrated four-point function via exchange Witten diagrams.} 

Before closing this section, let us note the following interesting fact. Suppose we add a quartic, non-derivative coupling $\lambda \phi^4$ to the free scalar theory. This lifts the flat direction associated to $\phi$. In the dual CFT, a non-vanishing $\beta$ function for the dual coupling $g$ is generated at two-loops, at leading order in $1/N$ (recall  that a one-loop coefficient $\beta_1$ cannot be generated by a quartic interaction at tree level in the bulk). 
In the bulk, the sign of $\lambda$ matters. In particular, the quartic interaction destabilizes the AdS background for $\lambda<0$, while it leaves AdS as a stationary point for $\lambda>0$. One can then try to understand what this instability corresponds to, in the dual CFT. The two-loop coefficient of the $\beta$ function in CPT is proportional to the (integrated) contact Witten diagram of figure \ref{Wdt4quartic}, which in this case is non-vanishing, {\it i.e.} $\beta_2= a \lambda$, with $a$ a positive $d$-dependent number, $a= \p^d\G(d/2)^4/2 \G(d)^3$. Therefore, $\beta_2$ has the same sign as $\lambda$. This means that for $\lambda>0$ the operator ${\cal O}$ becomes marginally irrelevant, while for $\lambda<0$ it becomes marginally relevant. Hence, in the latter case, a deformation triggered by ${\cal O}$ induces an RG-flow  which brings the theory away from the fixed point. On the contrary, for $\lambda>0$ the deformation is marginally irrelevant and the undeformed CFT remains, consistently, a stable point. Note how different this is from the case of SCFTs. There, marginal operators may either remain marginal or become marginally irrelevant, but never marginally relevant \cite{Green:2010da}, which agrees with the fact that AdS backgrounds are stable in supersymmetric setups.

\subsubsection{Loops in AdS}
\label{loopssec}

An obvious question is whether one can push the above analysis to higher orders in $1/N$. This corresponds to take into account loop corrections in the bulk. Already at one-loop, this is something very hard to do (see, {\it e.g.}, \cite{Cornalba:2007zb,Penedones:2010ue,Fitzpatrick:2011hu}, and, more recently, \cite{Aharony:2016dwx,Giombi:2017hpr}, where interesting progress have been obtained from complementary perspectives).  

\begin{figure}[ht]
\begin{center}
\includegraphics[height=0.25\textheight]{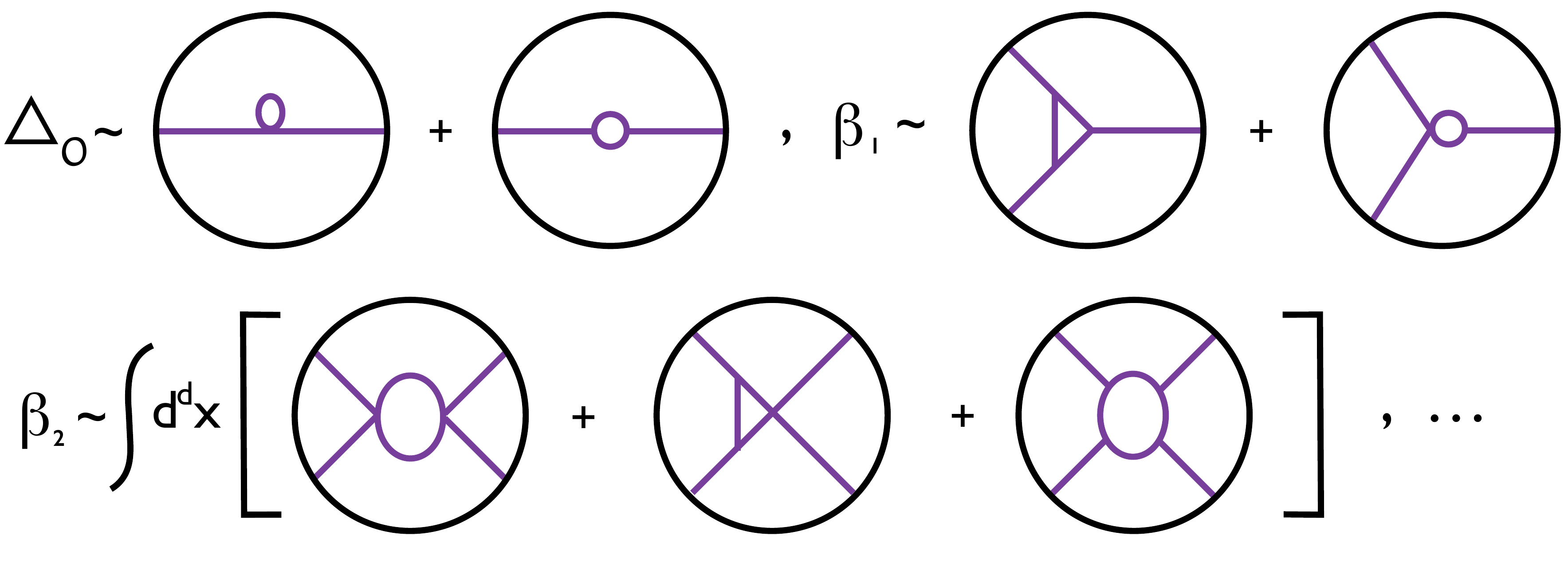}
\caption{One-loop Witten diagrams contributing to $\Delta_{\cal O}$, $\b_1$ and $\b_2$. Cubic and quartic Witten diagrams should include also those with loop corrections to propagators, but we have not drawn them explicitly.}
\label{Wdl}
\end{center}
\end{figure}

The main issue in this matter is not really to compute loop amplitudes per s\'e, but to make their relation to tree-level amplitudes precise, and this is something non-trivial to do in AdS. In fact, the question we are mostly interested in, here, is slightly different. Starting from the effective action \eqref{bulkgen1}, which is valid up to some energy cut-off $E$, in computing quantum corrections we are not much interested on how the couplings run with the scale but else on which (new) operators would be generated at energies lower than $E$.\footnote{In doing so, we can use the intuition from flat-space physics, since we are dealing with local effects in the bulk.} More precisely, what we have to do is to pinpoint, between the operators having passed our tree-level bulk analysis, {\it i.e.}, operators of the form \eqref{genop1}, those which could induce, at loop level, effective couplings which have instead been excluded at tree-level, that is the pure non-derivative couplings $\f^3$ and $\f^4$, as well as a mass term, which was set to zero from the outset. Such operators would spoil the vanishing of the $\b$ function, see figure \ref{Wdl} (the generation of a $\f^2$-term would modify the scaling dimension of ${\cal O}$, which should instead remain a marginal operator). 
So, the basic question we have to answer is whether (one and higher) loop analysis still leaves some of the operators \eqref{genop1} being compatible with the vanishing of the $\b$ function \eqref{betafull} and with $\Delta_{\Oc}=d$. 

That this is not an empty set can be easily seen as follows. Out of the full set \eqref{genop1}, let us consider shift-symmetric operators, only, namely operators which are invariant under the shift symmetry
\be
\label{shift1}
\phi \rightarrow \phi+a~.
\ee 
In perturbation theory, such operators cannot generate effective operators not respecting \eqref{shift1}, hence in particular $\f^n$ terms. Therefore, at least perturbatively, a conformal manifold does persist, if only shift-symmetric couplings are allowed in the action \eqref{bulkgen1}.

Let us now consider all other couplings, those with at least one naked $\f$, which do not respect the shift-symmetry \eqref{shift1}. Generically, these operators would generate any effective operator of the form $\f^n$, quantum mechanically. In particular, regardless of spacetime dimension, $\f^2$ and $\f^3$ will be generated at one-loop by any (non shift-symmetric) operator of the form \eqref{genop1}. Operators $\f^n$ with $n \geq 4$, instead, will be generated at one-loop or higher, depending on spacetime dimension and the specific operator \eqref{genop1} one is considering. 
In any event, the upshot is that, unless one invokes some unnatural tuning between the a priori independent couplings $\l_n$, any operator with at least one naked $\f$ should be excluded, eventually, by requiring a conformal manifold to persist at finite $N$. This leaves only shift-symmetric couplings in business, meaning that the shift symmetry \eqref{shift1} should be imposed on the bulk action \eqref{bulkgen1} altogether.\footnote{Following the discussion in the previous section, one can easily get convinced that the inclusion of a dynamical graviton, hence of the energy-momentum tensor in the low-dimension CFT operators, would not affect this result.}

As already noticed, shift-symmetric couplings would not contribute to (integrated) Witten diagrams not just at one loop but at any loop order in the bulk. Therefore, the final answer we got may be extended as a statement on the existence of a conformal manifold generated by ${\cal O}$ at all orders in the $1/N$ perturbative expansion. 

This apparently strong statement is just due to the axion-like behavior of an operator subject to eq.~\eqref{shift1}, which, as such, is expected to be lifted by non-perturbative effects only. The latter are suppressed as, say, $e^{-N}$. 
Richer holographic models would behave differently, and not share such perturbative non-renormalization property. Our analysis just aims at showing that, in principle, non-supersymmetric conformal manifolds can exist also beyond planar limit.  
It would be very interesting to consider models with richer structure. We will offer a few more comments on this issue in the next, concluding section.

\section{Discussion}

In general, it is hard  to find non-supersymmetric interacting CFTs in $d>2$, notable exceptions being, {\it e.g.}, the 3d Ising model, the critical $O(N)$ model and Banks-Zaks fixed point.\footnote{In the context of boundary conformal field theories (bCFT) there also exist examples. One such example, the mixed dimensional QED discussed in \cite{Teber:2012de,Herzog:2017xha}, is even believed to admit a (perturbative-stable) conformal manifold. We thank Chris Herzog for making us aware of this possibility.}  Since its early days, the AdS/CFT correspondence has been a natural framework where to look for novel examples.  Besides the limiting case of generalized free fields, most attempts have encountered obstructions.

Starting from the original ${\cal N}=4$ SYM/$\mbox{AdS}_5\times S^5$ duality, a very natural possibility is to consider non-supersymmetric orbifold thereof. It was shown in \cite{Dymarsky:2005uh} that (unlike in the parent supersymmetric theory) conformal invariance is broken already at leading order in $1/N$, by the logarithmic running of double-trace operators. This looks like a generic phenomenon which has been proposed in \cite{Dymarsky:2005nc,Pomoni:2008de} to be related to the presence of tachyonic instabilities in the gravity dual \cite{Adams:2001jb}.\footnote{Eventually, these problems might also be connected with recent claims about the non-perturbative instability of non-supersymmetric AdS vacua \cite{Ooguri:2016pdq}.} In this context, the only model we are aware of which evades this problem, is a non-tachyonic orientifold of Type 0B string theory, discussed in \cite{Angelantonj:1999qg}. However, it turns out that the absence of tachyons is not dual to the existence of fixed points in the dangerous double-trace operator running, but rather to the absence of such operators, at least at leading order in $1/N$ \cite{Liendo:2011da}. Hence, conformal invariance is preserved (and a fixed line exists in the space of couplings) but in a rather trivial sense, because of the exact equivalence of this theory with a subsector of the original ${\cal N}=4$ SYM, at large $N$.

More recently, another class of non-supersymmetric models obtained as a suitable double scaling limit of $\gamma$-deformed ${\cal N}=4$ SYM has been proposed \cite{Gurdogan:2015csr} (see also \cite{Sieg:2016vap}). For example, there exists a four-dimensional two (complex) scalar theory which looks particularly simple at face value. These models, although not being unitary, are interesting in many respect, but they also share the presence of double-trace operators in the effective action which spoil conformal invariance at leading order in $1/N$. In a more recent work \cite{Mamroud:2017uyz}, it was suggested that a suitable refinement of these models (that is, introducing an extra flavor structure for the component fields) could project out the double-trace operators, at least at leading order in $1/N$, similarly to \cite{Liendo:2011da}. And that also three and six-dimensional versions of the same model are not plagued by double-trace operator running, at large $N$. It would be interesting to see whether conformal invariance is preserved beyond leading order and, if this is the case, if a conformal manifold exists. These models look tractable enough, with respect to full-fledged top-down models, to make one hope that some concrete progress could be possible. More difficult, here, is to have some intuition about what the gravity dual description could be.

Within less ambitious, bottom-up models one can try to consider simple improvements of our one-field model. The basic reason why supersymmetric theories can admit conformal manifolds is due to the knowledge of the (perturbatively exact) $\b$ function for elementary fields and the possibility that some linear combinations have vanishing anomalous dimension $\g$, that is
\be
\label{an1}
\gamma_j(g_1,\dots, g_n)=0~~,~~j=1,2,\dots,m~,
\ee
where $g_i$ are the couplings associated to classically marginal operators ${\cal O}_i$. If $n>m$, the above equations describe a $n-m$ dimensional manifold of exactly marginal deformations, the conformal manifold. In a non-supersymmetric, bottom-up context, one can imagine  to deform a CFT by (say) two scalar marginal operators ${\cal O}_i$ as
\be
\delta S = g_1 \int d^d x \,{\cal O}_1 +  g_2 \int d^d x \, {\cal O}_2 ~,
\ee
with the couplings subject to the constrain 
\be
\label{cm2fields}
F(g_1,g_2)=0~.
\ee
This equation defines a line in the space of couplings. One can demand that on \eqref{cm2fields}, at one and two loops in CPT, $\b$ functions vanish and, generalizing the analysis of section \ref{cpt},  read-off the corresponding constraints that the existence of such one-dimensional conformal manifold imposes on the original CFT. Two-field models like the one above can be cooked-up holographically, and one might hope to get some richer answers with respect to the one-field model we have considered here.

\section*{Acknowledgements}

We would like to thank Riccardo Argurio, Agnese Bissi, Lorenzo Di Pietro, Denis Karateev, Per Kraus, Petr Kravchuk, Zohar Komargodski, Ioannis Papadimitriou, Rodolfo Russo, Slava Rychkov, Marco Serone, River Snively, Massimo Taronna, Giovanni Villadoro and Alexander Zhiboedov for helpful discussions and comments at different stages of this work. We are indebted with Riccardo Argurio, Lorenzo Di Pietro and Rodolfo Russo for very useful feedbacks on a preliminary draft version. H. R. is grateful to the  Department of Physics and Astronomy of UCLA for kind hospitality during the completion of this work. We acknowledge partial support from the MIUR-PRIN Project ``Non-perturbative Aspects Of Gauge Theories And Strings'' (Contract 2015MP2CX4).


\end{document}